\documentclass[aps,twocolumn]{revtex4-1}

\usepackage{color}
\usepackage{amsmath}

\usepackage[pdftex]{graphicx}
\DeclareGraphicsExtensions{.pdf,.jpeg,.png}

\usepackage{booktabs}
\usepackage{array,longtable}

\begin{document}

\title{Design and Simulation of Molecular Nonvolatile Single-Electron Resistive Switches}
\author{Nikita Simonian}
	\email{simonian@grad.physics.sunysb.edu}
\author{Konstantin K. Likharev}
	 \affiliation{Department of Physics and Astronomy, Stony Brook University, Stony Brook, NY 11794}
\author{Andreas Mayr}
\affiliation{Department of Chemistry, Stony Brook University, Stony Brook, NY 11794}
\date{\today}

\begin{abstract}
We have carried out a preliminary design and simulation of a single-electron resistive switch based on a system of two linear, parallel, electrostatically-coupled molecules: one implementing a single-electron transistor and another serving as a single-electron trap. To verify our design, we have performed a theoretical analysis of this “memristive” device, based on a combination of \textit{ab-initio} calculations of the electronic structures of the molecules and the general theory of single-electron tunneling in systems with discrete energy spectra. Our results show that such molecular assemblies, with a length below 10 nm and a footprint area of about 5 nm$^2$, may combine sub-second switching times with multi-year retention times and high ($> 10^3$) ON/OFF current ratios, at room temperature. Moreover, Monte Carlo simulations of self-assembled monolayers (SAM) based on such molecular assemblies have shown that such monolayers may also be used as resistive switches, with comparable characteristics and, in addition, be highly tolerant to defects and stray offset charges.
\end{abstract}

\maketitle


\section{Introduction}

Recently, a substantial progress was made in the fabrication of two-terminal ``memristive'' devices (including bistable ``resistive'' or ``latching'' switches) based on metal oxide thin-films, whose bistability is apparently based on the reversible formation/dissolution of conducting filaments --- see, e.g., recent reviews \cite{likharev08,waser09,StrukovKohlstedt12}. However, scaling of resistive memories and hybrid CMOS/nano-crossbar integrated circuits \cite{likharev08}, based on such switches, beyond the 10 nm frontier may still require more reproducible devices based on other physical principles. One possibility here is to use a molecular version of single-electron switches \cite{folling01}. 

Such a switch, schematically shown in Fig. 1a, is a combination of two electrostatically-coupled devices: a ``single-electron trap'' and a ``single-electron transistor'' \cite{likharev99} placed in parallel between two electrodes. (It will be convenient for us to call these electrodes the ``drain/control'' and the ``source'' --- see Fig. 1a.) When the charge state of the trap island is electroneutral ($Q = 0$), the Coulomb blockade threshold $V_C$ of the transistor is large (Fig. 1b), so that at all applied voltages with $|V| < V_C$ the transistor carries virtually no current --- the so-called OFF state of the switch. As soon as the voltage exceeds a certain threshold value $V_\leftarrow < V_C$, the rate of tunneling into the single-electron trap island increases sharply (Fig. 1c), and an additional elementary charge $q$ (either a hole or an electron) enters the trap island from the source electrode, charging it to $Q = q = \pm e$. The electrostatic field of this charge shifts the background electrostatic potential of transistor's island and as a result reduces the Coulomb blockade threshold of the transistor to a lower value $V_C^\prime$. This is the ON state of the switch, with a substantial average current flowing through the transistor at $V > V_C^\prime$. The device may be switched back into the OFF state by applying a reverse voltage in excess of the trap-discharging threshold $|V_\rightarrow|$. As was experimentally demonstrated for metallic, low-temperature prototypes of the single-electron switch \cite{dresselhaus94}, its retention time may be very long \footnote{A resistive switch with a sufficiently long (a-few-year) retention time at $V = 0$ may be classified as a nonvolatile memory cell. }.  However, for that the scale $e^2/2C$ of the single-electron charging energy of the trap island, with effective capacitance $C$, has to be much higher than the scale of thermal fluctuations, $k_{\mbox{\scriptsize B}}T$. For room temperature, this means the need for few-nm-sized islands \cite{likharev99}, and so far the only way of reproducible fabrication of features so small has been the chemical synthesis of suitable molecules --- see, e.g., \cite{tour03, cuniberti05}. 

Transport properties of single molecules, captured between two metallic electrodes, have been repeatedly studied by several research groups for more than a decade --- see, e.g., \cite{BummArnoldCygan96,Reed97,KerguerisBourgoinPalacin99,ParkPasupathyGoldsmith02,KubatkinDanilovHjort03,DanilovKubatkinKafanov08,WangBatsanovBryce09}. The practical use of such devices in VLSI circuits is still impeded by the unacceptably low yield of their fabrication and large device-to-device variability. The main reason of this problem is apparently the lack of atomic-scale control of the contacts between the molecules and the metallic electrodes. In addition, in three-terminal single-molecule devices that can work as transistors \cite{ParkPasupathyGoldsmith02,KubatkinDanilovHjort03}, there is an additional huge challenge of reproducible patterning of three-electrode geometries with the required sub-nm precision. These two challenges make single-molecule three-terminal stand-alone devices rather unlikely candidates for post-CMOS VLSI circuit technology. However, we believe that for resistive memories and CMOS/nano hybrids based on nano-crossbars (Fig. 1d \cite{likharev08}), these challenges may be met. Indeed, such crossbars use two-terminal crosspoint devices, so that their only critical dimension (the distance between the two electrodes) may be precisely controlled by layer thickness. In addition, if such devices are based on self-assembled monolayers (SAMs), the large number $N$ of molecules in a single device may mitigate the negative effects of interfacial and other uncertainties of single molecules \cite{akkerman07}. (Since the electrode footprint of a quasi-linear functional molecule stretched between the electrodes may be very small, $N$ may be as high as $\sim 10^2$ even for sub-10-nm-scale devices.)

The goal of this paper is to describe the results of the design and \textit{ab-initio} calculations of basic properties of a molecular resistive switch and SAMs based on such molecules. Our basic design is described, and its physics is discussed in the next section (Sec. II). In Sec. III we formulate a theoretical model which allows an approximate but reasonably accurate numerical simulation of electron transport properties of this device. The results of simulation of our most promising switch version are described in Sec. IV. In Sec. V we describe our approach to simulation of SAM layers consisting of resistive switch assemblies, and the results of these simulations. Finally, in Sec. VI we summarize our results and discuss the necessary further work towards the practical implementation of reproducible resistive switches.

\begin{figure}[!th]
\centering
\includegraphics[width=21pc]{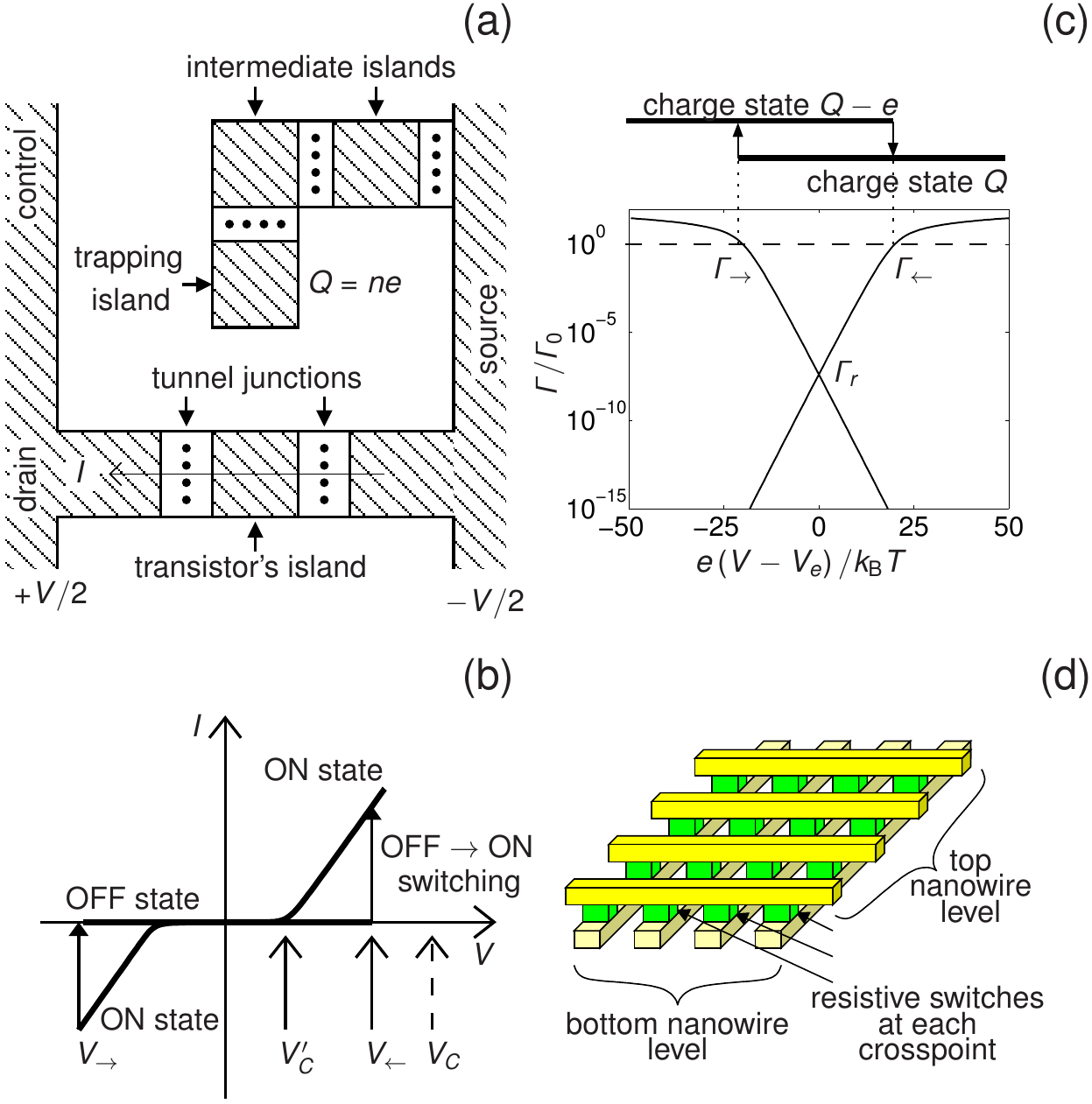}
\caption{(a) The traditional version of the single-electron resistive switch, (b) its $I-V$ curve (schematically), and (c) the ON/OFF switching rates of the device, calculated using the orthodox theory of single-electron tunneling \cite{averin91}, for $e^2/C = 20k_{\mbox{\scriptsize B}}T$. The inset in panel (c) schematically shows the switching between charge states of the trap resulting from repeated voltage sweeps with a rate $\Gamma_0 = \left|dV/dt\right|/(e/C) \ll \Gamma_r$. (d) Nano-crossbar with resistive switches as crosspoint devices.}
\label{fig1}
\end{figure}

\section{Resistive switch design}

Our initial design \cite{likharev03, mayr07} of the molecular resistive switch was based on oligophenyleneethynylene (OPE) chains as tunnel barriers and diimide (namely, pyromellitdiimide, naphthalenediimide, and perylenediimide) groups as trap and transistor islands. However, already the first quantitative simulations have shown that the relatively narrow HOMO-LUMO bandgap of the OPE chains (of the order of 1.5 eV \cite{ke07}) cannot provide a tunnel barrier high enough to ensure sufficiently long electron retention times in traps with acceptable lengths. As a result, we have concluded that alkane chains (CH$_2$-CH$_2$-...), with a bandgap of $\sim 9$ eV \cite{wangreed03}, are much better candidates for resistive switches. There is also substantial experience in the chemical synthesis of molecular electronic devices and SAMs using such chains as tunnel barriers \cite{tour03,cuniberti05}.

Figure 2 shows a possible realization of such a device, based on benzene-benzobisoxazole and naphthalenediimide acceptor groups (playing the role of single-electron islands), and alkane chains. In contrast with the usual (``orthodox'') design of the trap \cite{folling01,dresselhaus94,likharev89}, where a long charge retention time is achieved by incorporation of several additional single-electron islands into the trap charging path (Fig. 1a), the long alkane chain used in the molecular trap has a band structure which enables its use simultaneously in two roles: as a tunnel barrier as well as a replacement of intermediate islands.

\begin{figure}[!th]
\centering
\includegraphics[width=21pc]{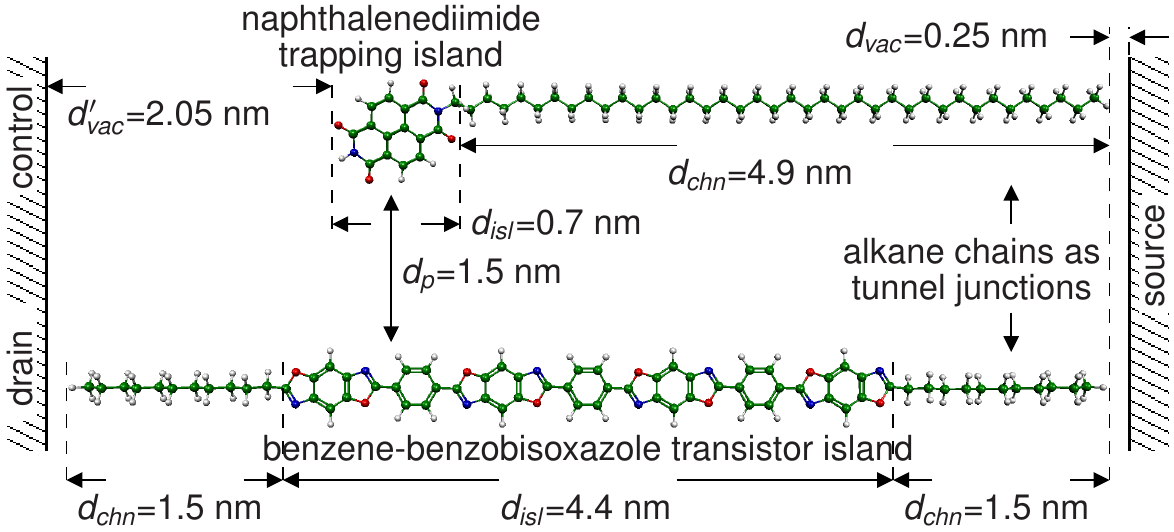}
\caption{Our final version of a molecular resistive switch featuring an alkane-naphthalenediimide single-electron trap electrostatically coupled to an alkane-benzobisoxazole-benzene single-electron transistor.}
\label{fig2}
\end{figure}

In order to explain this novel approach, let us first review the role of intermediate islands in the conventional design of the trap (Fig. 1a). If a single-electron island is so large that the electron motion quantization inside it is negligible, its energy spectrum, at a fixed net charge $Q$, may be treated as a continuum.  Elementary charging of the island with either an additional electron or an additional hole raises all energies in the spectrum by $e^2/2C$, where $C$ is the effective capacitance of the island \cite{likharev89,likharev99}. As a result, the continua of the effective single-particle energies of the system for electrons and holes are separated by an effective energy gap $e^2/C$ --- essentially, the ``Coulomb gap'' \cite{EfrosShklovskii75}. If this gap is much larger than $k_{\mbox{\scriptsize B}}T$ at applied voltages $V$ close to the  ``energy-equilibrating'' voltage $V_e$ (see the middle panel of Fig. 3a), it may ensure a very low rate $\Gamma_r$ of single-charge tunneling in either direction and hence a sufficiently long retention time $t_r = 1/\Gamma_r$ of the trap. The energy gap may be suppressed by applying sufficiently high voltages $V \sim e/C$ of the proper polarity, enabling fast switching of the device into the counterpart charge state --- see the left and right panels of Fig. 3a, and also Fig. 1c.

\begin{figure}[!th]
\centering
\includegraphics[width=21pc]{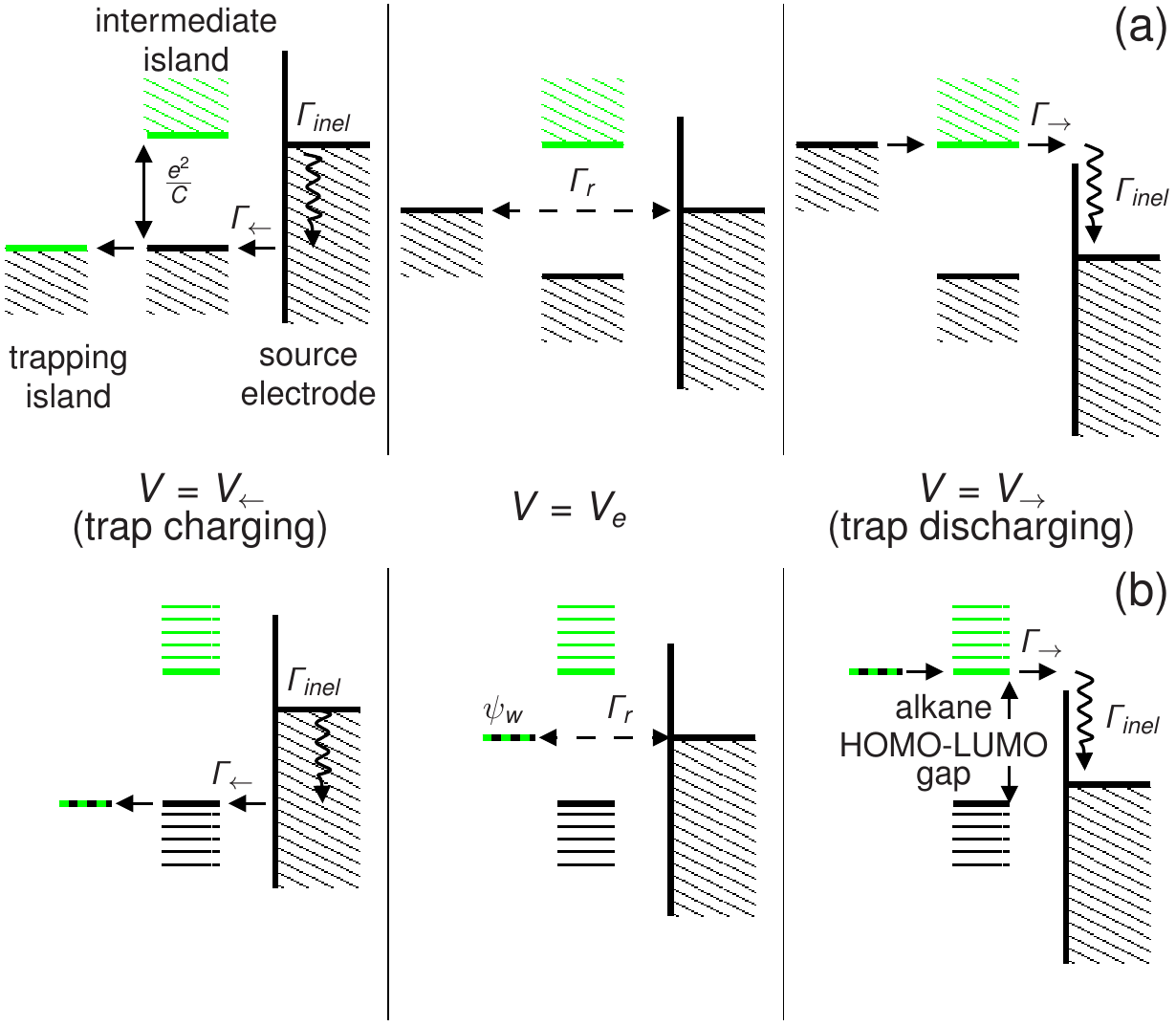}
\caption{Schematic single-particle energy diagrams of (a) the usual single-electron trap shown in Fig. 1a (for the sake of simplicity, with just one intermediate island) and (b) the molecular trap analyzed in this work (Fig. 2), each for three values of the applied voltage $V$. Occupied/unoccupied energy levels are shown in black/green. (The dotted green/black line denotes the energy level of the ``working'' orbital that is either empty or occupied during the device operation, defining its ON/OFF state.) Horizontal arrows show (elastic) tunneling transitions, while vertical arrows indicate inelastic relaxation transitions within an island, a molecule, or an electrode.}
\label{fig3}
\end{figure}

In the molecular single-electron trap shown in Fig. 2, the ``energy-equilibrating'' voltage $V_e$ aligns the Fermi energy of the source electrode with the lowest unoccupied level of the acceptor group that is, by design, located in the middle of the HOMO-LUMO gap of the alkane chain. As a result, an electron from the source electrode may elastically tunnel into the group only with a very low rate $\Gamma_r$ --- see the middle panel in Fig. 3b. The reciprocal process (at the same voltage) may be viewed as electron tunneling from the highest occupied molecular orbital of the singly-negatively charged molecule. (To simplify the terminology, in the reminder of the paper we call this molecular orbital the ``working orbital'' (indexed $W$), instead of HOMO or LUMO of the counterpart molecular ions, to make the name independent of the charge state of the device.) The energy-balance condition of both processes is similar, and may be expressed via the effective single-particle energy $\varepsilon_W$ of the working orbital \cite{simonian07}:

\begin{equation}
\varepsilon_W=\Delta E(n) \equiv E_{gr}(n)-E_{gr}(n-1),
\label{eq_epsilon_def_ground}
\end{equation}
where $E_{gr}(n)$ is the ground-state energy of the molecular ion with $n$ electrons. (In the case of singly-negative ion we are discussing, $n = n_0 + 1$ \footnote{We use the notation in which the fundamental electric charge unit $e$ is positive, so that the electric charge of an ion with $n$ electrons is $Q(n) = -e(n-n_0)$.}, where $n_0$ is the number of protons in the molecule.) In this notation, the energy-balance (level-alignment) condition, which defines the voltage $V_e$, is

\begin{equation}
\varepsilon_W=W-e\gamma V_e,
\label{eq_epsilon_balance_ground}
\end{equation}
where $W$ is the workfunction of the source electrode material, and $\gamma$ is a constant factor imposed by the geometry of the junction; $0 < \gamma  < 1$. (Its physical meaning is the fraction of the applied voltage, which drops between the trapping island and the source electrode.) At the charging threshold voltage $V_\leftarrow$, energy $\varepsilon_W$ becomes aligned with the valence band edge of the chain, allowing for a fast charging of the molecule --- see the left panel in Fig. 3b. Similarly, as shown on the right panel in Fig. 3b, at $V_{\rightarrow}$ this energy becomes aligned with the conduction band edge of the chain, allowing for a fast discharging of the molecule.

As an example, Fig. 4a shows the atomic self-interaction corrected (ASIC) \cite{pemmaraju07} Kohn-Sham electron eigenenergy spectrum $\varepsilon_{i}^{\mbox{\scriptsize ASIC}}(n)$ of the alkane-naphthalenediimide molecule used as our final trap design (Fig. 2), with the net charge $Q(n) = -e (n-n_0) = -e$, as a function of the applied voltage $V$. (Here $i$ is the spin-orbital index; see Sec. III below.) Point colors in Fig. 4a crudely represent the spatial localization of the orbitals: blue corresponds to their localization at the trapping (acceptor) group, while red marks the localization at the alkane chain's part close to the source electrode. Figure 4b shows the probability density of the working orbital $\psi_W^{\mbox{\scriptsize ASIC}}=\psi_{n_0+1}^{\mbox{\scriptsize ASIC}}(n_0+1)$  of the molecular trap, integrated over the directions perpendicular to the molecule's axis, with blue lines corresponding to the probability density at the most negative applied voltage. At the equilibrating voltage $V_e \approx 2.2 $ V, the working orbital is well localized at the acceptor group, and is isolated from the source electrode by a $\sim 4.5$-eV-high energy barrier. However, as Fig. 4b shows, the probability density of the orbital decays into the alkane group rather slowly, with the exponent coefficient $\beta \approx 0.4a_{\mbox{\scriptsize B}}^{-1}$, corresponding (in the parabolic approximation of the dispersion relation) to the effective electron mass $m_{ef} \approx 0.1 m_0$ \footnote{Experiments (for a recent summary, see, e.g., Table 1 in \cite{mcdermott09}) give for the exponent coefficient $\beta$ a wide range $(0.26-0.53) a_{\mbox{\scriptsize B}}^{-1}$ corresponding to the effective mass range $(0.05-0.2) m_0$ (assuming a rectangular, 4.5-eV-high energy barrier). It has been suggested \cite{mcdermott09} that such a large variation is due to a complex dispersion law inside the alkane bandgap, making $\beta$ a strong function of the tunneling electron energy.}.  As a result, a long ($\sim 5$ nm) alkane chain is needed to ensure an acceptable retention time of the trap. (The 2-nm free-space separation between the other side of the molecule and the control/drain electrode, shown in Fig. 2, is quite sufficient for preventing electron escape in that direction.)

\begin{figure}[!th]
\centering
\includegraphics[width=21pc]{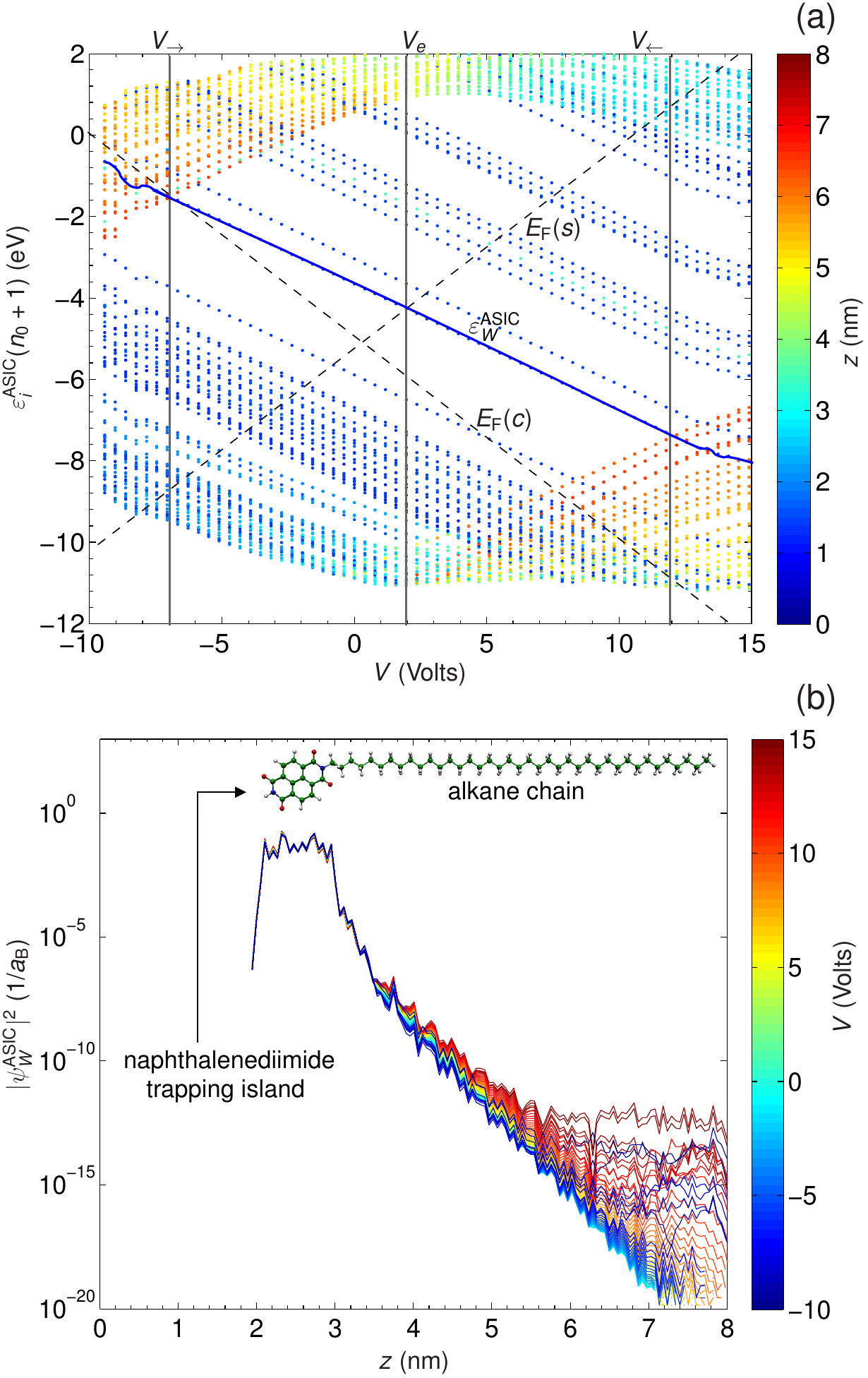}
\caption{ASIC density-functional-theory (DFT) results (corrected for level ``freezing'' at high positive and negative voltages --- see Sec. III for details) for the singly-negatively charged alkane-naphthalenediimide trap molecule. (a) Kohn-Sham energy spectrum as a function of the applied voltage $V$, with colors representing the spatial localization of the corresponding orbitals  --- see the legend bar on the right. The vertical lines mark voltage values $V_\leftarrow$, $V_e$, and $V_\rightarrow$ corresponding to the left, middle, and right panels of Fig. 3b. The dashed lines labeled $E_{\mbox{\scriptsize F}}(s)$ and $E_{\mbox{\scriptsize F}}(c)$ show the Fermi levels of the source and control/drain electrodes whose workfunction was assumed to equal 5 eV. (b) The ``working'' orbital's probability density, integrated over the directions perpendicular to molecule's axis, for a series of applied voltages $V$ --- see the legend bar on the right of the panel.}
\label{fig4}
\end{figure}

At a sufficiently high forward/reverse bias voltage, the working orbital energy $\varepsilon_W$ crosses into the conduction/valence band of the alkane chain, so that the orbital partly hybridizes with the states localized near the source electrode interface, described by the rise of $|\psi_W|^2$ at larger values of $z$ --- see Fig. \ref{fig4}b. This rise enables fast electron tunneling to/from the source electrode, i.e. a fast switching of the device to the counterpart charge state, in a manner similar to that of the conventional single-electron trap, as shown schematically on the left and right panels of Fig. \ref{fig3}b. Thus the long molecular chain, with a sufficiently large HOMO-LUMO gap, may indeed play the roles of both the tunnel junction and intermediate islands of the ``orthodox'' single-electron trap.

For the design of the second component of the switch, the molecular single-electron transistor, the most important challenge is to satisfy the ON and OFF state current requirements. In particular, the ON current should not be too large to keep the power dissipation in the circuit at a manageable level, but simultaneously not too small, so that the device output does not vanish in the noise of the sense amplifier (for memory applications \cite{StrukovLikharevMemory05,strukov07}) or the CMOS invertor (in hybrid logic circuits \cite{StrukovLikharevLogic05}). Also, the ON/OFF current ratio should be sufficiently high to suppress current ``sneak paths'' in large crossbar arrays \cite{franzon05, strukov07}.  In addition, the transistor molecule should be geometrically and chemically compatible with the trap molecule, enabling their chemical assembly as a unimolecular device, with their single-electron island groups sufficiently close to provide substantial electrostatic coupling. (Without it, the charge of the trap would not provide a substantial modulation of the transistor current.) At the same time, the molecules must not be too close, in order to prevent a parasitic discharge of the trap via electron cotunneling through the transistor into one of the electrodes. The chemical compatibility strongly favors the use of similar chains as the transistor's tunnel junctions.

\begin{figure}[!th]
\centering
\includegraphics[width=21pc]{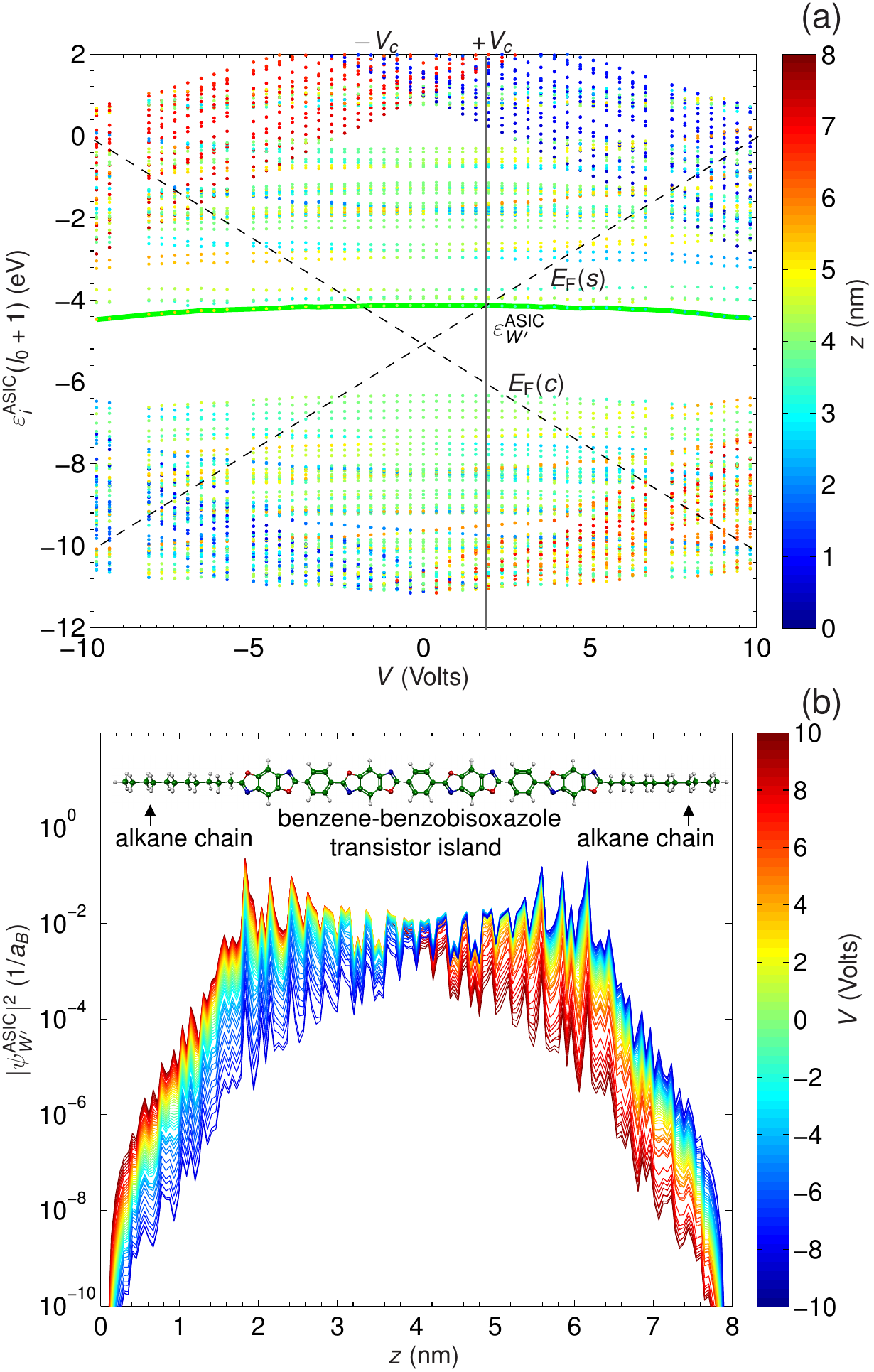}
\caption{ASIC results for the single-negatively charged benzene-benzobisoxazole transistor molecule. (a) Kohn-Sham energy spectrum as a function of the applied voltage $V$, with colors representing the spatial localization (within the junction) of the corresponding orbitals  --- see the legend bar on the right. The dashed lines labelled $E_{\mbox{\scriptsize F}}(s)$ and $E_{\mbox{\scriptsize F}}(c)$ show the Fermi levels of the source and control/drain electrodes whose workfunction was assumed to equal 5 eV. (b) Probability density of the working orbital, integrated over the directions perpendicular to the molecular axis, for a series of applied voltages --- see the legend bar on the right.}
\label{fig5}
\end{figure}

We have analyzed several alkane-chain based transistor devices with naphthalenediimide, perylenediimide and benzobisoxazole acceptor groups as transistor islands. However, in all these cases the long alkane chains, needed to match the lengths of the transistor and trap molecules, make ON currents too low. Finally, we have decided to use an unusually long ($\sim 4.3$-nm) benzene-benzobisoxazole \cite{krausea88} island group --- see Fig. \ref{fig2} and Fig. \ref{fig5}b. Figure \ref{fig5}a shows the Kohn-Sham electron eigenenergy spectrum $\varepsilon_{i}^{\mbox{\scriptsize ASIC}}(l_0+1)$ of this molecule as a function of voltage $V$ (where $l_0$ is the number of protons in the transistor molecule). Blue/red colored points correspond to the orbitals localized at the left/right alkane chain, while green color points denote the orbitals extended over the whole acceptor group. This extension is clearly visible in Fig. 5b, which shows the probability density of the working orbital $\varepsilon_{W^\prime}^{\mbox{\scriptsize ASIC}}=\varepsilon_{l_0+1}^{\mbox{\scriptsize ASIC}}(l_0+1)$ of the transistor molecule. During transistor operation, the tunneling electron may populate any of several group-localized orbitals, resembling the operation of the usual (metallic) single-electron transistor. As a result of such island extension, alkane chains of the transistor could be substantially shortened, to $\sim 1.5$-nm-long C$_{11}$H$_{25}$, enabling low but still acceptable ON currents of the order of 0.1 pA, even if a small (0.25-nm) vacuum gap between the alkane chain and the source electrode is included into calculations to give a phenomenological description of the experimentally observed current reduction due to unknown interfacial chemistry \cite{simonian07}.

\section{Theoretical model and approximations}

Each molecule used in our device has a discrete set of possible excited states, and hence the electron transport is not limited to a single channel. In order to take into account all of these channels, we have used the ``quasi-single-particle approximation'' whose simplest version had been first formulated by Averin and Korotkov for semiconductor quantum dots \cite{AverinKorotkov90, averin91} and which was recently generalized \cite{simonian07} to be more applicable to molecular structures. In this approximation, the energy of an arbitrary state $k=\{n,i\}$ of the molecule equals

\begin{equation}
E_k=E_{gr}(n)+\sum_{i>n}\varepsilon_i(n)p_i-\sum_{i \leq n}\varepsilon_i(n)(1-p_i),
\label{eq_eex_def}
\end{equation}
where coefficients $\varepsilon_i(n)$ have the physical meaning of single-particle excitation energies of an $n$-electron ion, and numbers $p_i$ (equal to either 0 or 1) are the single-particle energy level occupancies. The condition of elastic tunneling, leading to a transition between states $k$ and $k^\prime$, is given by the natural generalization of Eqs. (\ref{eq_epsilon_def_ground}) and (\ref{eq_epsilon_balance_ground}):

\begin{equation}
\varepsilon_{k \rightarrow k^\prime} = W - e\gamma V,
\label{eq_epsilon_balance}
\end{equation}
where the single-electron recharging/excitation energy is now defined as
\begin{equation}
\varepsilon_{k \rightarrow k^\prime} \equiv E_k - E_{k^\prime}.
\label{eq_epsilon_def}
\end{equation}

Because of the large size and complexity of the molecules used in our design, the only practical way to calculate their electronic structure is to use a software package (such as SIESTA \cite{soler02}) \footnote{Initially, we made an attempt to use NRLMOL \cite{pederson00} which had been successfully employed in our previous study of single-electron tunneling through smaller molecules \cite{simonian07}. However, we have found the performance of SIESTA (with the ``standard'' double-Zeta polarized basis set) for our current problem to be substantially higher, though the results obtained from NRLMOL may be slightly more accurate.}, based on the density-functional-theory (DFT) \cite{jones89}, which may provide a reasonably accurate ground state energy $E_{gr}^{\mbox{\scriptsize DFT}}(n)$ and a single-particle spectrum $\varepsilon_{i}^{\mbox{\scriptsize DFT}}(n)$ at a fraction of the computational cost of more correct \textit{ab-initio} methods. Unfortunately, for such strongly correlated electronic systems as molecules considered in this paper, results obtained using standard DFT software packages \footnote{this is valid not only for the DFT packages based on the local spin density approximation (LSDA), such as the standard version of SIESTA. Another popular DFT functional, the generalized gradient approximation (GGA) \cite{PerdewBurkeErnzerhof96}, does not provide much improvement on these results} have significant self-interaction errors \cite{pedrew81}.   

We believe the source of such errors is that the approximate treatment of the exchange-correlation term in the Kohn-Sham Hamiltonian does not completely cancel the self-interaction energy present in the ``Hartree term'' of the  Hamiltonian \footnote{In contrast, in the Hartree-Fock theory the exchange energy is exact (of course, in the usual sense of the first approximation of the perturbation theory), and the self-interaction errors are absent \cite{pedrew81}.}.  Indeed, the standard DFT approach leads to errors, in the key energies (\ref{eq_epsilon_def_ground}) and (\ref{eq_epsilon_def}), of the order of the single-electron charging energy $e^2/2C$, where $C$ is the effective capacitance of the island group --- see Appendix A for details.  This error may be rather substantial; for example in the naphthalenediimide-based trap molecule (Fig. 2), it is approximately equal to 1.8 eV. For this reason, the electron affinity $E_{gr}(n_0+1)-E_{gr}(n_0)$, calculated using the LSDA DFT for the singly-negatively charged ion of the molecular trap, is significantly (by $\sim 3.2$ eV) larger than the experimental value of similar molecules \cite{bhozale08, singh06}. The LSDA energies may be readily corrected to yield a much better agreement with experiments (see Table 1 in Appendix A), however, it is not quite clear how such a theory may be used for a self-consistent calculation of the corresponding working orbital $\psi_W(\mathbf{r})$. 

We have found that a significant improvement may be achieved by using the recently proposed Atomic Self-Interaction Corrected DFT scheme (dubbed ASIC \cite{pemmaraju07}) implemented in a custom version of the SIESTA software package. For the molecules that we have considered here, this approach gives the Kohn-Sham energy $\varepsilon_{n_0+1}(n_0+1)$ very close to the experimental electron affinity. However, we have found that using even this advanced approach for our task faces two challenges. 

First, the algorithm gives (at least for our molecular trap states with $n = n_0+1$ and $n = n_0 + 2$ electrons) substantial deviations from the relation $\varepsilon_W = \varepsilon_{n+1}(n+1)$ for $n = n_0$ (which has to be satisfied in any exact theory \cite{janak78, pemmaraju07}), with the ground energy difference (\ref{eq_epsilon_def_ground}) close to the LSDA DFT results. This means that Eq. (\ref{eq_epsilon_def}) cannot be directly used with the ASIC results; instead, for the electron transfer energy between adjacent ions $n$ and $n - 1$ we have used the following expression:

\begin{equation}
\varepsilon_{k \rightarrow k^\prime} = \varepsilon_{i^\prime}^{\mbox{\scriptsize ASIC}}(n).
\label{eq_epsilon_asic}
\end{equation}

This relation implies that the differences $\varepsilon_{i^\prime}^{\mbox{\scriptsize ASIC}}(n)-\varepsilon_{n}^{\mbox{\scriptsize ASIC}}(n)$ describe all possible single-particle excitations within the acceptor group, if the index $i^\prime$ is restricted to orbitals localized on the group. (Other orbitals, localized on the alkane chain are irrelevant for our calculations since they do not contribute to the elastic tunneling between the molecular group and the electrode.)

In order to appreciate the second problem, look at Fig. 6 which shows the voltage-dependent Kohn-Sham spectra of the singly-negatively charged molecular trap, calculated using the ASIC SIESTA package for $T > 0$ K. Notice that above voltage $V_t \approx 13$V, and below voltage $V_t^\prime \approx -7$V, the eigenenergy spectrum is virtually ``frozen''. (The LSDA SIESTA gives similar results.). As explained in Appendix B using a simple but reasonable model (similar to that used in Appendix A), at $V > V_t$ such ``freezing'' originates from the spurious self-interaction of an electron whose wavefunction cloud is gradually shifted from the top occupied orbital of the valence band of the chain, with energy $\varepsilon_v$, into the initially empty group-localized orbital with energy $\varepsilon_{W+1}$. (A similar freeze at voltages $V < V_t^\prime$, is due to the spurious gradual transfer of the electron wavefunction cloud from the working orbital, localized at the acceptor group, with energy $\varepsilon_W$, to the lowest orbital of the conduction band of the chain, with energy $\varepsilon_c$.) It is somewhat surprising that this spurious effect (which should not be present in any consistent quantum-mechanical approach --- see Appendix B) is so strongly expressed in the ASIC version of the SIESTA code, which was purposely designed to get rid of the self-interaction in the first place. Being no SIESTA experts, we may only speculate that the nature of this artifact is related to the smoothing of the derivative discontinuity present in the ASIC method as the electron number passes through an integer value, which is mentioned in \cite{pemmaraju07} --- see also Fig. 7 in that paper. 

\begin{figure}[!th]
\centering
\includegraphics[width=21pc]{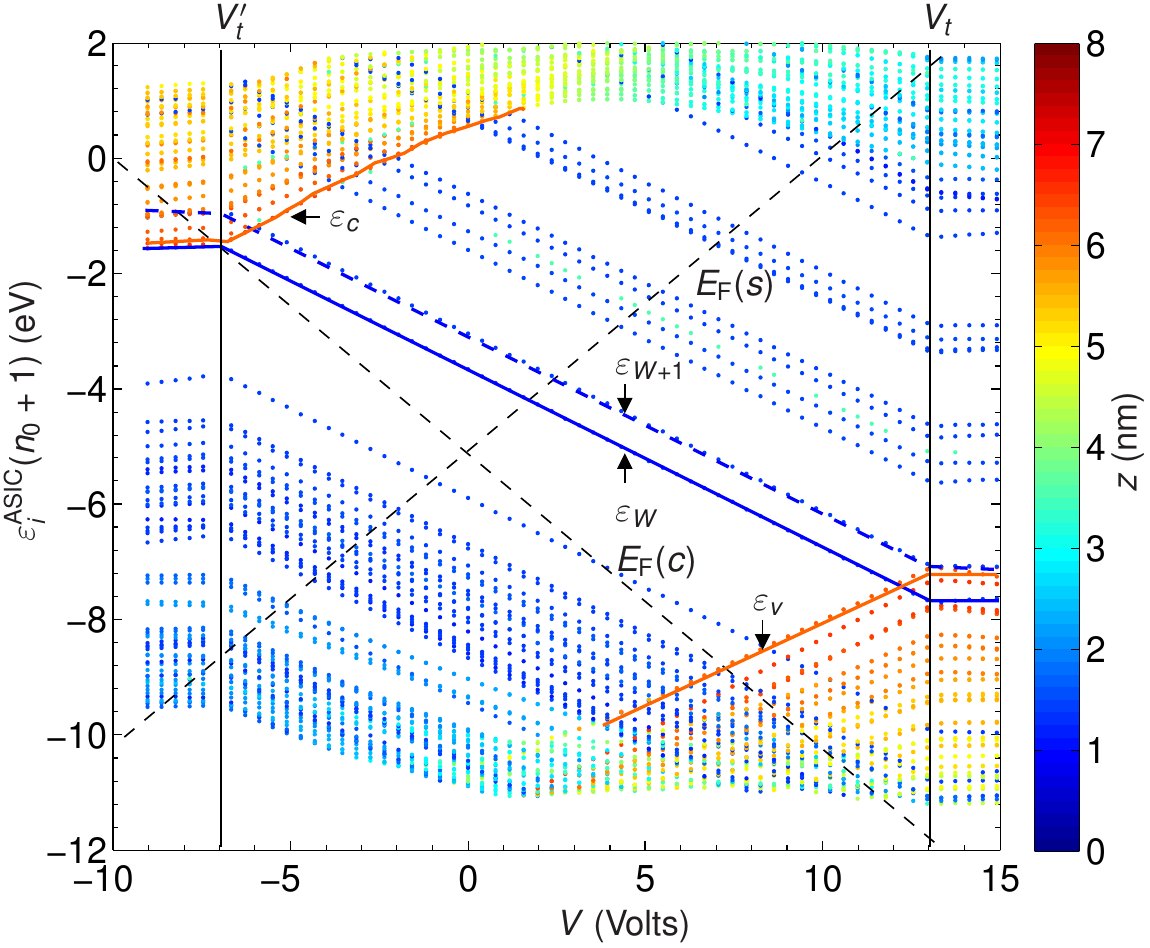}
\caption{The Kohn-Sham spectra of the singly-negatively charged molecular trap, calculated with the ASIC SIESTA at $T = 10$ K. At voltages below $V_t^\prime  \approx -7$ V, the spectrum is virtually frozen due to a spurious gradual shift of the highest-energy electron from the ``working'' orbital (with energy $\varepsilon_W$, shown with a solid blue line) localized on the acceptor group, to the lowest orbital (with energy $\varepsilon_c$, shown with a solid red line) of the conduction band of the alkane chain. As a result, the calculated spectrum is virtually voltage-insensitive (``frozen''). In the voltage range $V_t^\prime < V < V_t \approx 13$ V, ASIC SIESTA gives apparently correct solutions, with the working orbital $\varepsilon_W$ fully occupied, and the next group-localized orbital (with energy $\varepsilon_{W+1}$, the dashed blue line) completely unoccupied. However, at $V > V_t$ the package describes a similar spurious gradual shift of the highest-energy electron from the highest level $\varepsilon_v$ of the valence band of the chain to orbital $\varepsilon_{W+1}$, resulting in a similar spectrum ``freeze''. The spectrum evolution, calculated after the (approximate) correction of this spurious ``freezing'' effect, is shown in Fig. 4a above. }
\label{fig6}
\end{figure}

Fortunately, there is a way to correct this error very substantially by following the iterative process of self-consistent energy minimization within ASIC SIESTA. Indeed, for a fixed temperature $T > 0$ K (when the program automatically populates molecular orbitals in accordance with the single-particle Fermi-Dirac statistics) and voltages $V > V_t \approx 13$ V and $V < V_t^\prime \approx -7$ V, its iterative process converges to a wrong solution with the energy levels frozen at their $V_t$ and $V_t^\prime$ values, as is discussed above --- see Fig. 6. However, if the temperature in that program is fixed at $T = 0$ K, its iterative process ends up in quasi-periodic oscillations between different solutions --- most of them with frozen levels (just like in Fig. 6), but some of them with the group localized energies like the working orbital energies $\varepsilon_W$, $\varepsilon_{W+1}$ and the valence/conduction band edge energies $\varepsilon_v$, $\varepsilon_c$ close to their expected (unfrozen) values. (Those values were obtained by a linear extrapolation of their voltage behavior calculated at $V_t^\prime < V < V_t$.) Since such a solution is repeated almost exactly at each iterative cycle (see the vertical boxes in Fig. 7), we believe that it is close to the correct solution expected from the self-consistent quantum-mechanical theory --- see Appendix B. These approximate solutions were used in our calculations both above $V_t$ and below $V_t^\prime$; they are illustrated in Fig. 4a, where we have substituted the incorrect ``frozen'' solutions for $T > 0$ K with solutions for $T = 0$ K, with $\varepsilon_W  \approx \varepsilon_{W}^{\mbox{\scriptsize fit}}$, $\varepsilon_{W+1}  \approx \varepsilon_{W+1}^{\mbox{\scriptsize fit}}$ and $\varepsilon_{c}  \approx \varepsilon_{c}^{\mbox{\scriptsize fit}}$ at $V < V_t^\prime$ or $\varepsilon_{v}  \approx \varepsilon_{v}^{\mbox{\scriptsize fit}}$ at $V > V_t$. Let us emphasize that the approximate nature of these solutions may have affected our calculations (we believe, rather insignificantly), only at $V > V_t \approx 13$ V and $V < V_t^\prime \approx -7$ V, i.e. only the device recharging time results, but not the most important retention time calculations for smaller voltages --- see Fig. 9 below.

\begin{figure}[!th]
\centering
\includegraphics[width=21pc]{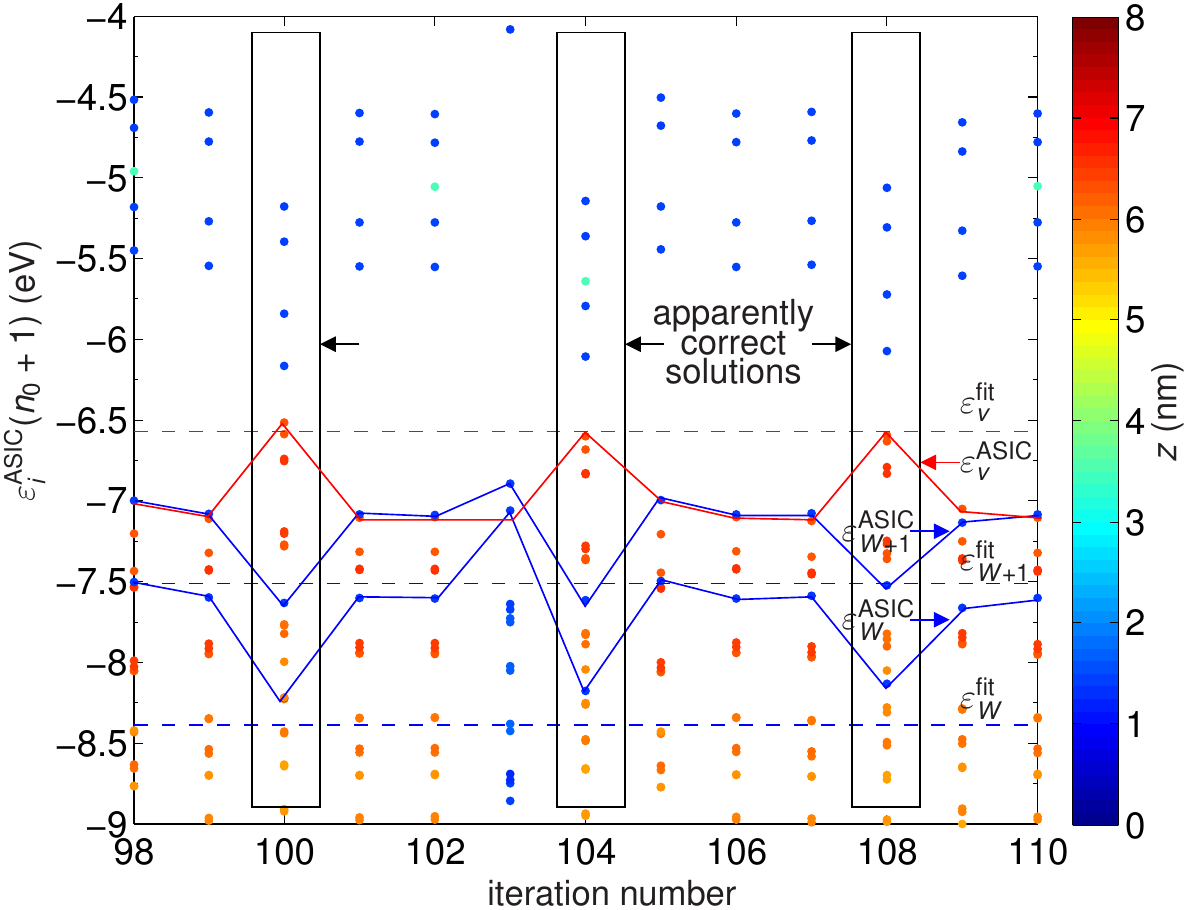}
\caption{The Kohn-Sham energy spectrum of our trap molecule, as calculated by successive iterations within ASIC SIESTA for $T = 0$ K and $V = 14.9$ V, i.e. above the threshold voltage $V_t \approx 13$ V. Vertical boxes mark the apparently correct solutions with energies of the working orbital ($\varepsilon_W^{\mbox{\scriptsize ASIC}}$), the next  group-localized orbital ($\varepsilon_{W+1}^{\mbox{\scriptsize ASIC}}$), and the highest orbital of the valence band of the alkane chain ($\varepsilon_v^{\mbox{\scriptsize ASIC}}$) all close to their respective values  $\varepsilon_W^{\mbox{\scriptsize fit}}$, $\varepsilon_{W+1}^{\mbox{\scriptsize fit}}$  and $\varepsilon_v^{\mbox{\scriptsize fit}}$ obtained by a linear extrapolation of their voltage dependence calculated at $V_t^\prime < V < V_t$. Just like in Figs. 4a, 5a and 6, point colors represent the spatial localization of the corresponding orbitals. Lines are only guides for the eye.}
\label{fig7}
\end{figure}

With the electron orbitals and eigenenergies calculated, we have described dynamics of both the trap and the transistor, just as in our first work \cite{simonian07}, by a set of master equations for state probabilities \cite{averin91}, which are valid because of the incoherent character of single-electron tunneling to/from the continuum of electronic states in metallic electrodes \cite{likharev99}. Moreover, for the inelastic relaxation rates $\Gamma_{\mbox{\scriptsize inel}}$ and the rates $\Gamma_\leftarrow$ and $\Gamma_\rightarrow$ of the elastic tunneling between the molecular group and the metallic electrodes (see arrows in Fig. 3), the following strong inequality, 

\begin{equation}
\Gamma_{\mbox{\scriptsize inel}} \gg \Gamma_\leftarrow, \Gamma_\rightarrow
\label{eq_gamma_ineq}
\end{equation}
is well fulfilled. (Indeed, the rates $\Gamma_{\mbox{\scriptsize inel}}$ are crudely of the order of $10^{12}$ 1/s in both molecules and metals. On the other hand, our results, described in Sec. IV below, yield transistor currents $I \sim 10^{-13}$ A, meaning that $\Gamma_\leftarrow$ and $\Gamma_\rightarrow$  are of the order of  $I/e  \sim 10^6$ 1/s in the transistor; the rates are even much lower than that in the trap --- see Fig. 9b.) Relation (\ref{eq_gamma_ineq}) allows us to account only for the tunneling events starting from thermal equilibrium, and ensures that the rates $\Gamma_{\mbox{\scriptsize inel}}$ drop out of the calculations. 

In comparison with \cite{simonian07}, one more new element of this work is the electrostatic coupling between the trap and the transistor which features similar but much more frequent single-charge transitions. This rate hierarchy allows the trap to be described by averaging rates $\Gamma$ of tunneling events in it over a time interval much longer than the average time period between tunneling events in the transistor, but still much shorter than $1/\Gamma$. These average rates may be calculated as

\begin{equation}
\left<\Gamma_{\rightarrow}\right> = \sum_{l}\sigma_{n_0}(l)w_{l,+}(n_0),
\label{eq_gammar}
\end{equation}
for electron tunneling from the source into the trap molecule, and

\begin{equation}
\left<\Gamma_{\leftarrow}\right> = \sum_{l}\sigma_{n_0+1}(l)w_{l,-}(n_0+1),
\label{eq_gammal}
\end{equation}
for the reciprocal event. Here $\sigma_n(l)$ are the conditional probabilities of certain charge states $l$ of the transistor island provided that the trap is in the $n$-electron charge state (with $n$ equal to either $n_0$ or $n_0+1$), while $w_{l,\pm}(n)$ are the total rates of single electron tunneling between the trap in its initial charge state $n$ and the source electrode. These rates have been calculated using Eq. (11) in \cite{simonian07}, with an extra index $l$ added to account for the transistor's state. The conditional probabilities $\sigma_n(l)$ satisfy the usual normalization condition:

\begin{equation}
\sum_l \sigma_n(l)=1,
\label{eq_sigmanorm}
\end{equation}
and (together with the dc current $I$ flowing through the transistor) have been calculated as in \cite{simonian07}, by combining the master equations of single-electronics \cite{AverinKorotkov90, averin91} with \textit{ab-initio} calculations of molecular orbitals and spectra, and the Bardeen formula \cite{bardeen61} for tunneling rates. 

\begin{figure}[!th]
\centering
\includegraphics[width=21pc]{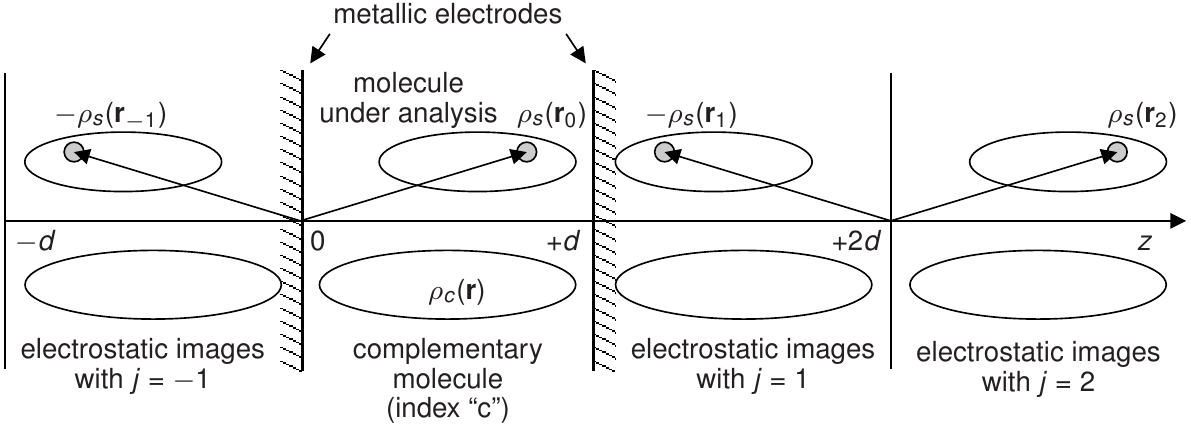}
\caption{A schematic view of charge densities participating in Eq. (\ref{eq_coulomb}).}
\label{fig8}
\end{figure}

\begin{figure}[!th]
\centering
\includegraphics[width=21pc]{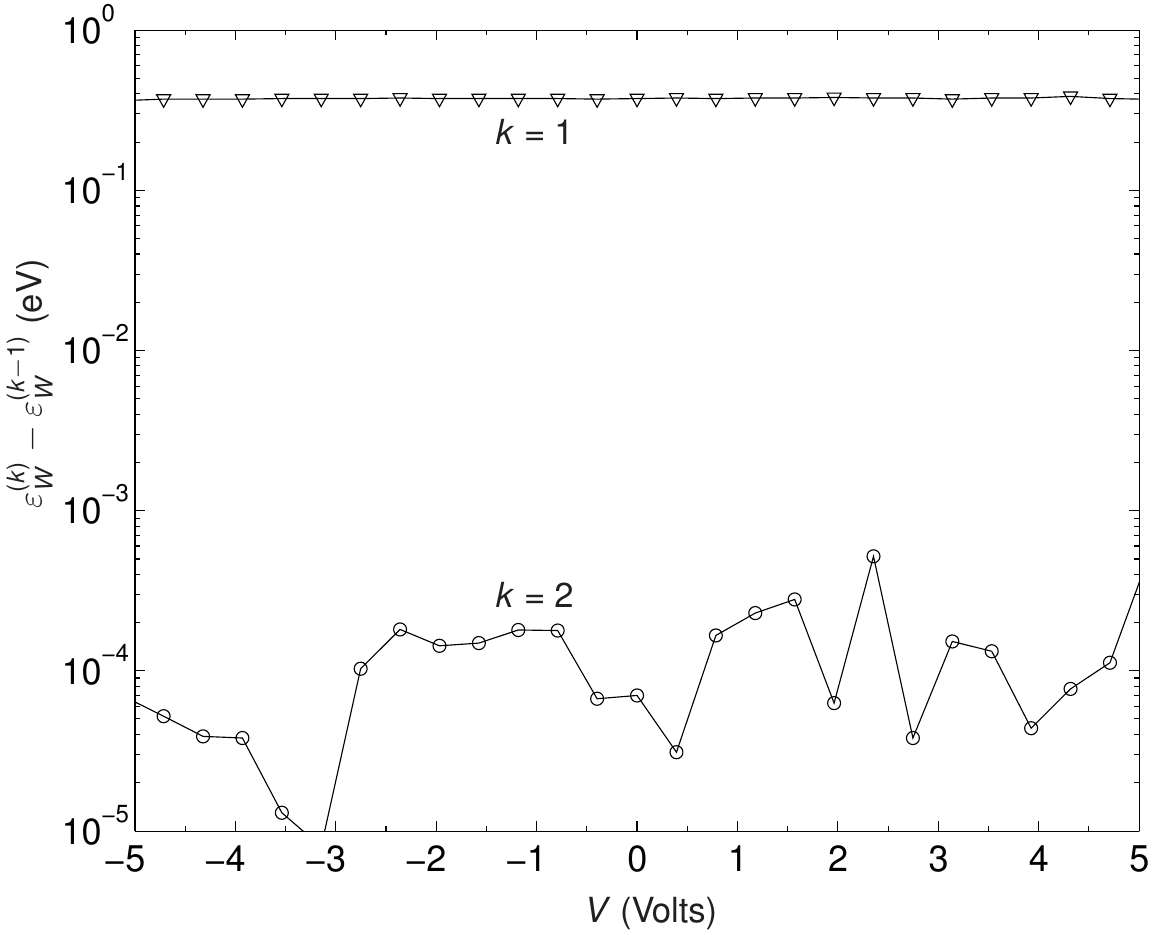}
\caption{Differences between the energies of the working orbital of the molecular trap, calculated with ASIC SIESTA at the $k$-th and $(k-1)$-th iterations of the Coulomb interaction potential, given by Eq. (\ref{eq_coulomb}), as functions of the applied voltage.}
\label{fig9}
\end{figure}

The electrostatic interaction between the two molecules is taken into account by an iterative incorporation of the numerically calculated Coulomb potential created by both molecules (as well as by the series of their charge images in the metallic electrodes of the system, which we have assumed to be plane, infinite surfaces --- see Fig. \ref{fig8})  into the Kohn-Sham potentials. From the elementary electrostatics, this potential may be expressed as

\begin{gather}
\begin{split}
\phi_s(\mathbf{r})=&\int\frac{\rho_{c}(\mathbf{r_0})}{\left|\mathbf{r}-\mathbf{r_0}\right|}d^3r_0\\
   &+\sum_{j\neq 0}(-1)^{j} \int \frac{\rho_{c}(\mathbf{r_j})+\rho_{s}(\mathbf{r_j})}{\left|\mathbf{r}-\mathbf{r_j}\right|}d^3r_j \\
   &-V\frac{z-d/2}{d},
\end{split} \nonumber \\
\mathbf{r}_j \equiv \mathbf{r}_0 + \mathbf{n}_z \times \left\{ \begin{split}
 jd, &\mbox{ for $j$ even,} \\
  (j+1)d-2z, &\mbox{ for $j$ odd.}
       \end{split} \right.
   \label{eq_coulomb}
\end{gather}
where $\rho_{s}(\mathbf{r}_0)$, $\rho_{c}(\mathbf{r}_0)$ are the total charge distributions of the molecule under analysis and the complementary molecule, and $\rho_{s}(\mathbf{r}_j)$, $\rho_{c}(\mathbf{r}_j)$ are the corresponding charge images in the source $(j>0)$ and the drain $(j<0)$ electrodes --- see Fig. \ref{fig8}. The first term in Eq. (\ref{eq_coulomb}) is the potential created by the complement molecule, the second term describes the potential of the infinite set of charge images of both molecules in the source and control/drain electrodes, and the third term is the potential created by the applied source-drain voltage. At the 0-th iteration, the first two terms are taken equal to zero. Fortunately, the iterations give rapidly converging results, so that there was actually no need to go beyond the second iteration --- see Fig. \ref{fig9}.

Another important change introduced into our calculations is that the charge transfer rates (see Fig. 4c in \cite{simonian07}) are calculated in a simpler way. Namely, one of the key conditions of validity of the Bardeen formula for the tunneling matrix elements

\begin{equation}
T_{[s,c],i}=\frac{\hbar^{2}}{2m}\int_{S}\left(\psi_{s,c}^{\ast}\frac{\partial \psi_i}{\partial z} - \psi_i^\ast\frac{\partial \psi_{s,c}}{\partial z}\right)dS
 \label{eq_bardeen}
\end{equation}
(where $\psi_i$  is the molecular orbital, $\psi_{s,c}$ are the wavefunctions of electrons located inside the source or control/drain electrodes, and $S$ is an arbitrary surface separating the single-electron island from the corresponding electrode) is that the result given by Eq. (\ref{eq_bardeen}) is independent of the position of the surface $S$. Due to electrostatic screening of the electric field by the electrode, the Kohn-Sham potential becomes very close to the vacuum potential at just a few Bohr radii $a_{\mbox{\scriptsize B}}$ away from the molecule's last atom. Therefore, if the surface $S$ is selected inside the vacuum gap between the molecule's end and the electrode surface, the effect of the molecule on wavefunctions $\psi_{s,c}$ is negligible with good accuracy (corresponding to a fraction of one order of magnitude in the resulting current). Hence, these wavefunctions may be calculated analytically to describe the usual exponential 1D decay into vacuum, instead of a numerical solution of a Schr\"odinger equation, as it had been done in \cite{simonian07}.

On top of the \textit{ab-initio} calculation scheme, we have performed the following check of the component molecule spacing.  As was mentioned in Sec. II, the molecules have to be placed sufficiently far from each other to prevent a parasitic discharge of the trap via the elastic cotunneling through the transistor island, into one of the electrodes. This effect may be estimated using the following formula \cite{stoof96,averin00}:

\begin{equation}
\Gamma_\rightarrow^{\mbox{\scriptsize cot}}=\frac{\Delta^2 w_{\rightarrow}}{\hbar^2 w_{\rightarrow}^2 + 2\Delta^2+4\varepsilon^2},
 \label{eq_averin}
\end{equation}
where $\Delta$ is the matrix element of electron tunneling between the trap and the transistor islands (that exponentially depends on the distance $d_p$ between the molecules, see Fig. \ref{fig2}), $\varepsilon$ is the difference between eigenenergies of these states, and $w_{\rightarrow}$ is the rate of tunneling between the transistor island and its electrodes. We have found that in order for the cotunneling rate $\Gamma_\rightarrow^{\mbox{\scriptsize cot}}$ to be below the retention rate of the trap $\Gamma_r$, the distance $d_p$ should not be lower than $\sim 1.5$ nm. Such relatively large separation justifies separate DFT calculations of the electronic structures of the trap and the transistor (with the molecular geometry of each component device initially relaxed, using LSDA SIESTA \footnote{The geometry relaxation was done for isolated neutral molecules only, at no applied bias voltage and without the account of the possible image charge effects; in the relaxed geometry all force components on the atoms are smaller than $0.05$ eV/$\AA$. To justify this procedure, we have verified that trap charging and discharging rates, calculated using the trap geometry relaxed in the presence of  a high (8 V) bias, do not significantly differ from the rates (shown in Fig. \ref{fig10}b) calculated for its relaxation at $V = 0$.}), related only by their electrostatic interaction described by Eq. (\ref{eq_coulomb}).

\section{Simulation results for a single resistive switch}

\begin{figure}[!th]
\centering
\includegraphics[width=21pc]{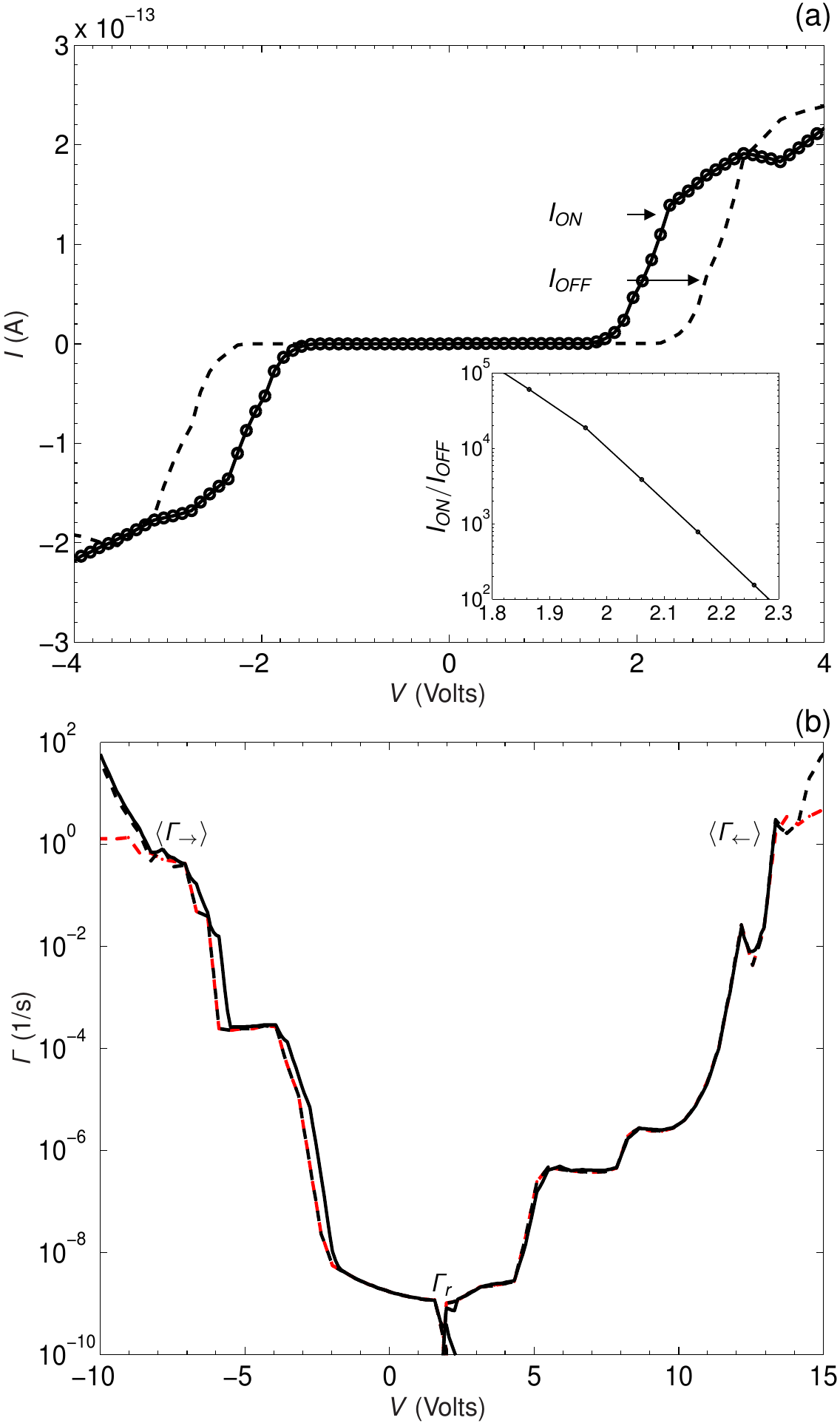}
\caption{(a) Calculated dc $I-V$ curves of the transistor for two possible charge states of the trap molecule. The inset shows the ON/OFF current ratio of the transistor on a semi-log scale, within the most important voltage interval. (b) The trap switching rates, calculated with (solid lines) and without (dashed lines) taking into account the transistor's back action, as functions of the applied voltage. Red dashed lines on panel (b) show the trap switching rates calculated without the level ``freezing'' correction (Sec. III and Appendix B) and without taking into the account the transistors' back action.}
\label{fig10}
\end{figure}

Figure \ref{fig10} shows our main results for the resistive switch shown in Fig. \ref{fig2}, for temperature $T = 300$ K. They include the dc $I-V$ curves of the transistor, plotted in Fig. \ref{fig10}a for both charge states of the trap, and the rates of transitions between the neutral and single-negatively charged states of the trap, with and without the account of the transistor effect on the trap molecule (Fig. \ref{fig10}b). The plots show that the resulting $I-V$ curves fit our initial specifications rather well, with a broad voltage window (from $\sim 2.0$ to $\sim 2.5$ V) for the trap state readout, a large ON/OFF current ratio within that window (inset in Fig. \ref{fig10}a), and the ON current $I \sim 0.2$ pA.

Figure \ref{fig10}b shows that the trap features a high retention time, $\tau_r  > 10^8$s, for both charge states, within a broad voltage range, $-2$V $< V <$ $+5$V. (It is somewhat surprising how little is the trap retention affected by the electrostatic ``shot noise'' generated by fast, quasi-periodic charging and discharging of the transistor island, which is taken into account by our theory.) The range includes point $V = 0$, so that the device may be considered a nonvolatile memory cell. 

At the same time, the device may be switched between its states relatively quickly by applied voltages outside of this window. The price being paid for using alkane chains with their large HOMO-LUMO gap is that the voltages necessary for fast switching are large --- they must align the valence or conduction band of the alkane chain with the group-localized working orbital --- see Fig. \ref{fig4}a.

\section{SAMs of resistive switches}

\begin{figure}[!th]
\centering
\includegraphics[width=21pc]{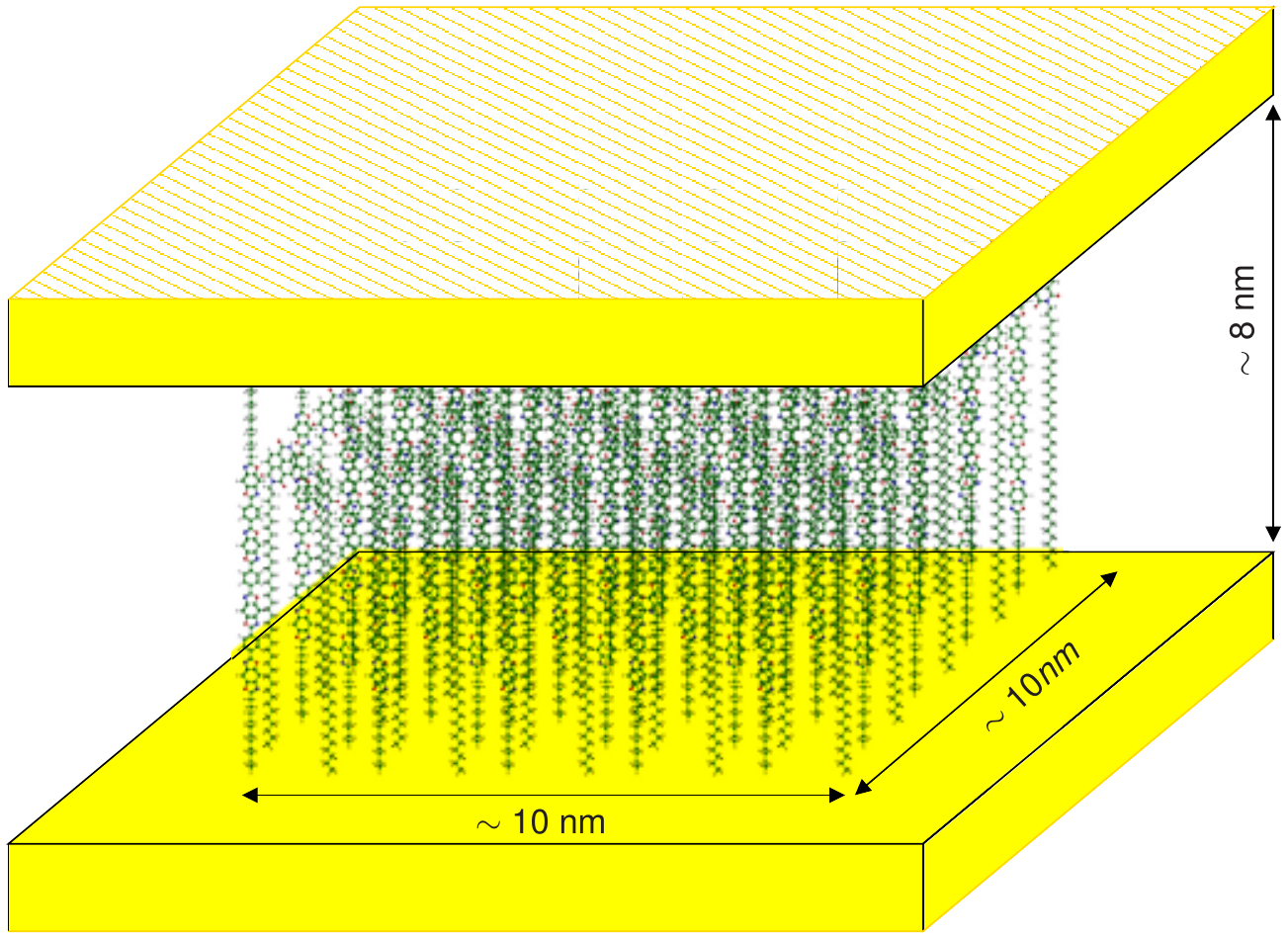}
\caption{Schematic view of a $5 \times 5$-switch SAM sandwiched between two electrodes.}
\label{fig11}
\end{figure}

Probably the largest problem of molecular electronics \cite{tour03,cuniberti05} is the low reproducibility of interfaces between molecules and metallic electrodes. However, recent results \cite{akkerman06} indicate that this challenge may be met at least for self-assembled monolayers (SAMs) encapsulated using special organic counter-electrodes. This is why we have explored properties of SAMs consisting of square arrays of $N \times N$ resistive switches described above --- see Fig. \ref{fig11}. In order to increase the tolerance of the resulting SAM devices to self-assembly defects and charged impurities, it is beneficial to place the component molecular assemblies (Fig. 2) as close to each other as possible, say at distances comparable to that ($\sim 1.5$ nm) between the trap and transistor. In this case, the Coulomb interactions between the component molecules are very substantial, and properties of the system have to be calculated taking these interactions into account.

A system of $N \times N$ resistive switches has $2N \times 2N$ single-electron islands and hence at least $2^{2N \times 2N}$ possible charge states, which would require solving that many master equations for their exact description. Even for relatively small $N$, this approach is impracticable, and virtually the only way to explore the properties of the system is to perform its Monte Carlo simulations \cite{likharev89, likharev99}. In this method a random number generator is used twice for each state change: first, to calculate the random time of some state change (which obeys Poissonian statistics), and second, to calculate the charge transition type (if several transitions are possible simultaneously). The procedure requires a prior calculation of rates of transitions between all pairs of charge configurations which differ by one single-electron tunneling event.

\begin{figure}[!th]
\centering
\includegraphics[width=21pc]{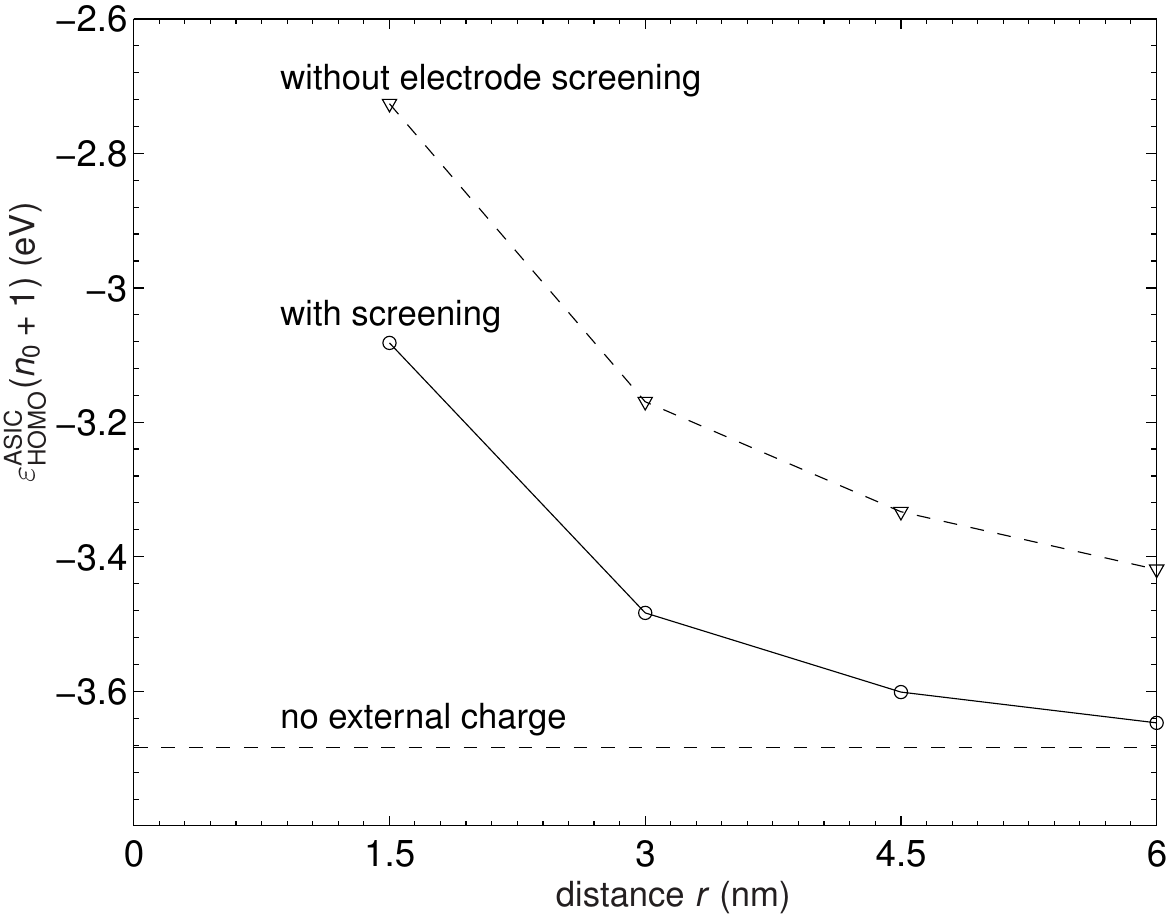}
\caption{Effect of a single charge of a trap molecule on the electron affinity of another molecule, located at distance $r$ without and with the account of the electric field screening by the common metallic electrodes.}
\label{fig12}
\end{figure}

As was discussed above, the peculiarity of our particular system is that it features two very different time scales: the first one (for our devices, $\tau_t \approx e/I_{ON} \sim 10^{-4} - 10^{-6}$ s) characterizes fast charge tunneling through single-electron transistors, and the second one corresponds to the lifetimes of trap states ($\tau_r = 1/\Gamma \sim 10^8 - 10^{-2}$ s). In order to gather reasonable statistics of the switching rates, our data accumulation time, for each parameter set, corresponded to the physical times of up to 10 s, i.e., included up to a million transition tunneling events in the system's transistors.

\begin{figure}[!th]
\centering
\includegraphics[width=21pc]{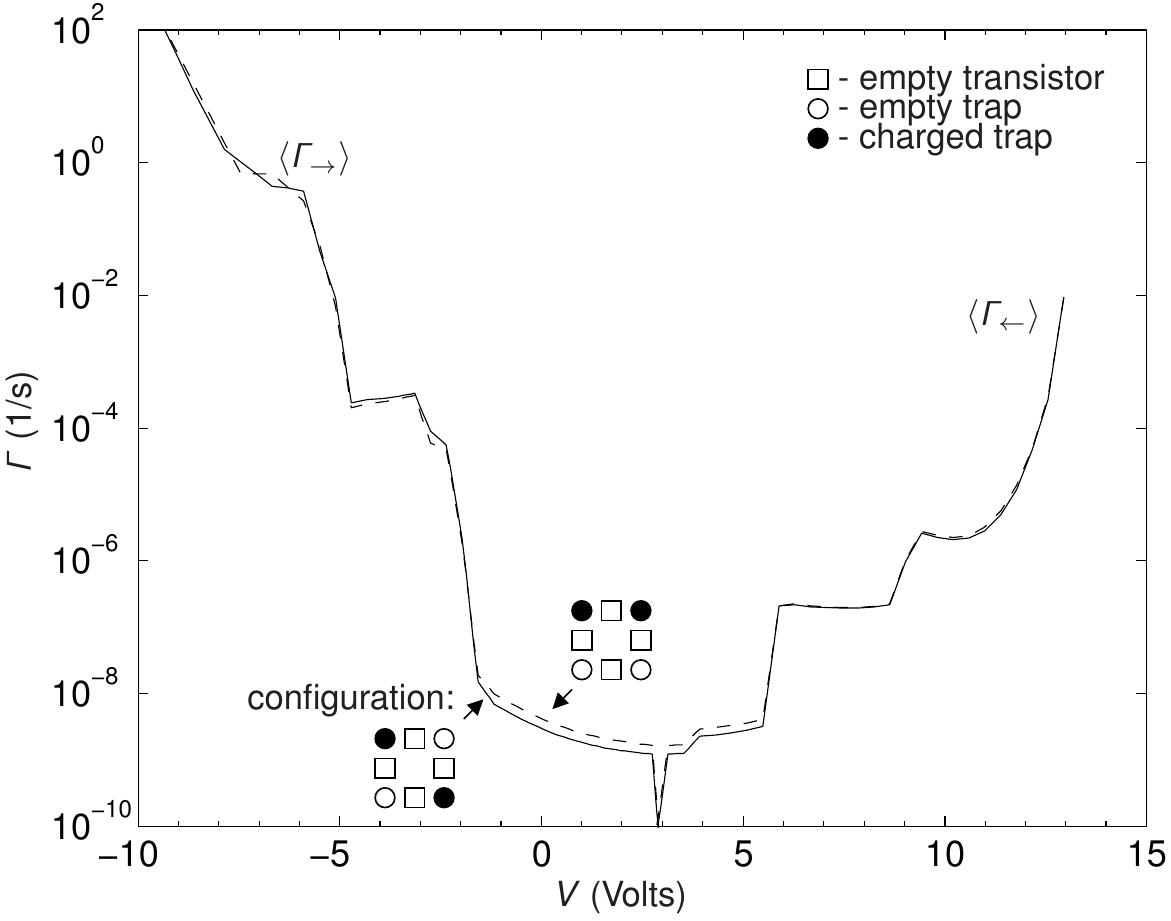}
\caption{Trap tunnel rates as functions of the applied voltage for two quasi-similar nearest-neighbor charge configurations shown in the insets.}
\label{fig13}
\end{figure}

As a check of the validity of the procedure, the Monte Carlo algorithm was first applied to a single resistive switch, and it indeed gave virtually the same result as the master equation solution. We then used the approach for a direct simulation of SAM fragments with two and more coupled resistive switches. As the fragment is increased beyond a $2 \times 2$ switch array, even the Monte Carlo method runs into computer limitations, because of the exponentially growing number of the possible charge configurations. The calculations may be very significantly sped up by using the approximation in which each molecule's state affects the potential of only its nearest neighbors. This approximation has turned out to be very reasonable (Fig. \ref{fig12}) and may be justified by the fact that metallic electrodes of the system substantially screen the Coulomb potential of the charges of distant molecules: the distance between the acceptor group centers and the electrodes, $d/2 \approx 4$ nm, is of the same order as the 3-nm distance between the molecule and its next-next neighbors. In this nearest-neighbor approximation, each molecule (a trap or a transistor) is still affected by 8 other molecules. To limit the number of the charge configurations even further, we have treated all ``essentially similar'' of them (having charge pairs at equal distances, irrespective of their angular position) as identical --- see Fig. \ref{fig13}.

\begin{figure}[!th]
\centering
\includegraphics[width=21pc]{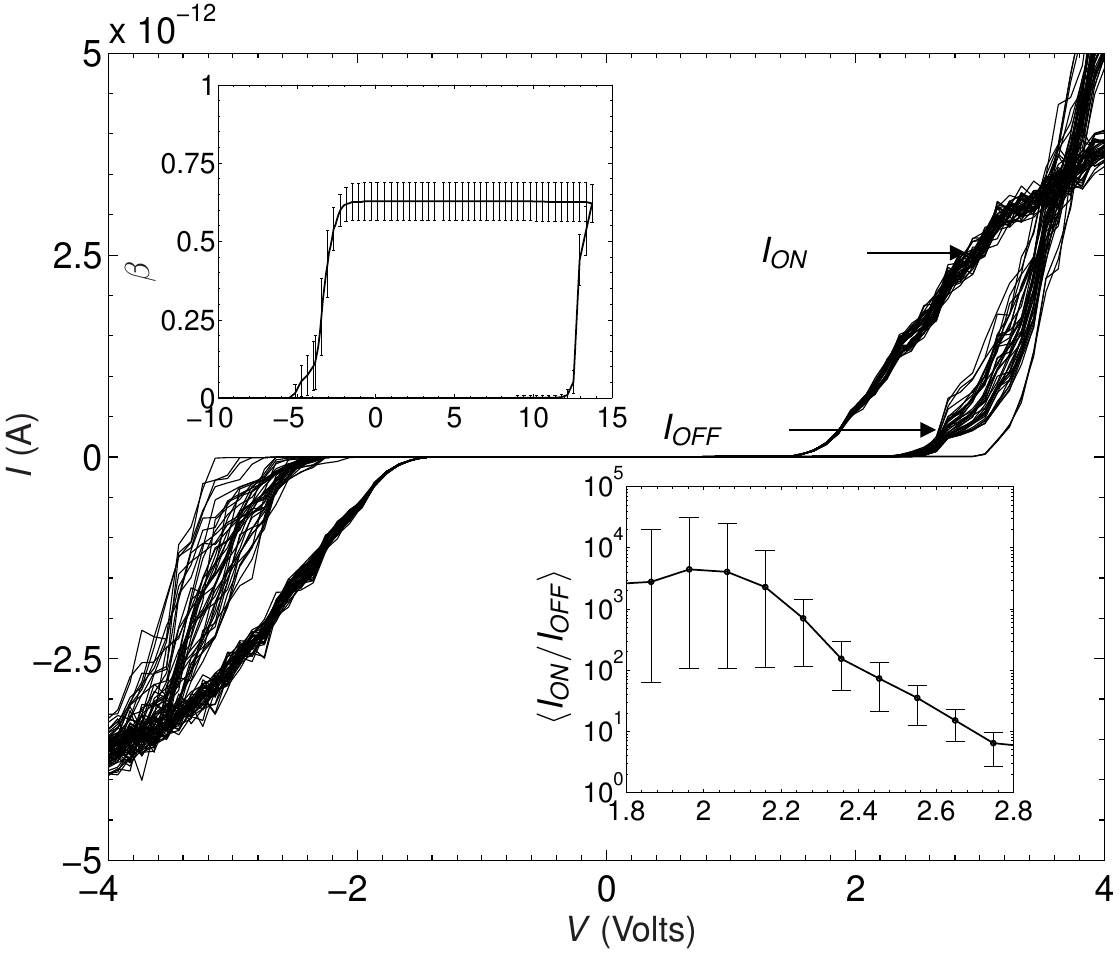}
\caption{Monte-Carlo simulated dc $I-V$ curves of a 25-switch SAM. The top inset shows the fraction $\beta$ of single-negatively charged traps, averaged over 40 sweeps of applied voltage between -8V and 13V. The bottom inset shows the ON/OFF current ratio averaged over the voltage sweeps, and its maximum sweep-to-sweep spread.}
\label{fig14}
\end{figure}

Figure \ref{fig14} shows the results of calculations, based on this approach, for a $5 \times 5$-switch SAM, of the total area close to $10 \times 10$ nm$^2$. The switching and state readout properties are very comparable with those of a single switch (Fig. \ref{fig10}), despite a significant mutual repulsion between single electrons charging neighboring traps. In order to better understand why this repulsion does not have adverse effects on the operation of the SAM as a whole, we have calculated the correlation coefficients of charging of two molecules in the SAM as a function of the distance between them. At voltages above the transistor Coulomb blockade, transistor molecules switch their charge state fast and the correlation coefficient $K(r)$ between two transistor molecules may be calculated directly from their time evolution records at a constant $V$. On the other hand, trap molecules have quasi-stationary charge states, so that the correlation between two trap molecules has to be calculated from a set of snapshots of their charge states (at some voltage of interest) taken at repeated, slow sweeping of the applied voltage throughout the whole voltage range. 

\begin{figure}[!th]
\centering
\includegraphics[width=21pc]{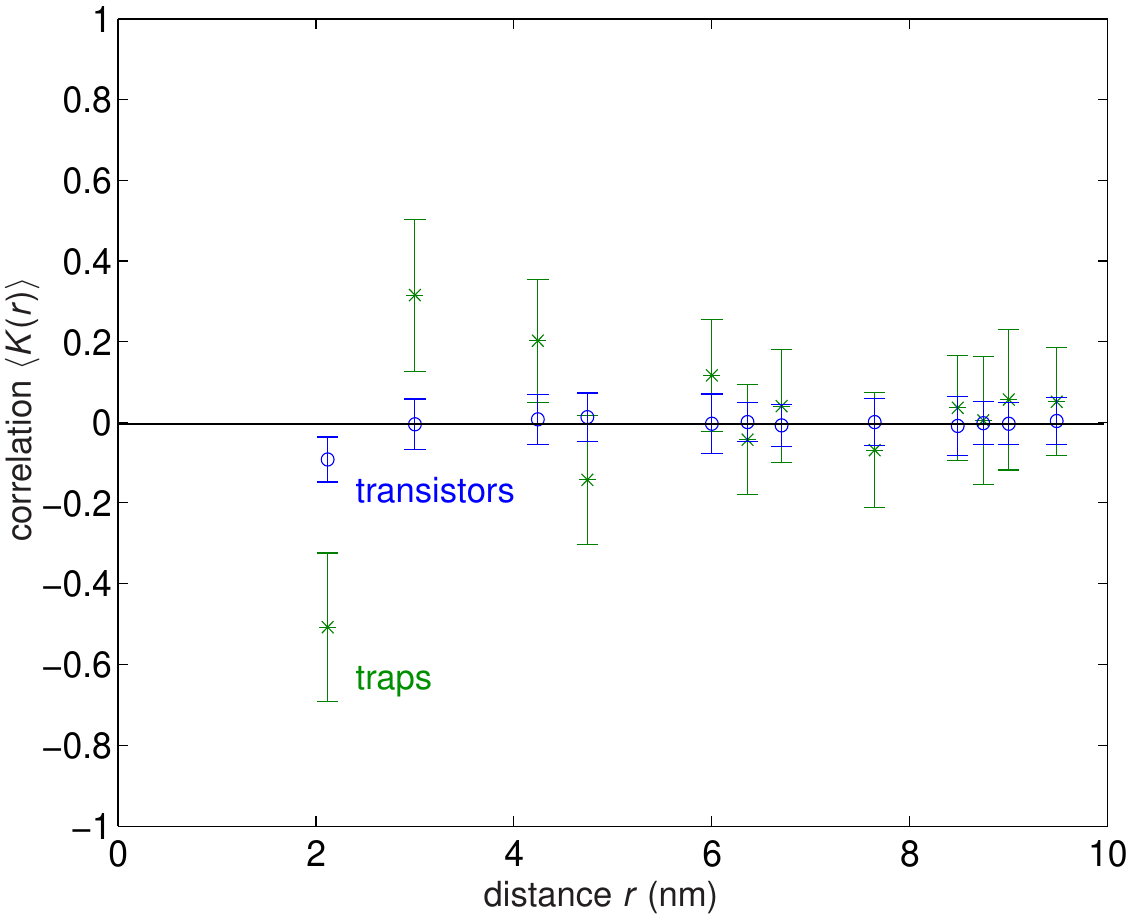}
\caption{The average correlation between two traps (green) and two transistors (blue) as a function of distance between them in a $5 \times 5$ device SAM.}
\label{fig15}
\end{figure}

\begin{figure}[!th]
\centering
\includegraphics[width=21pc]{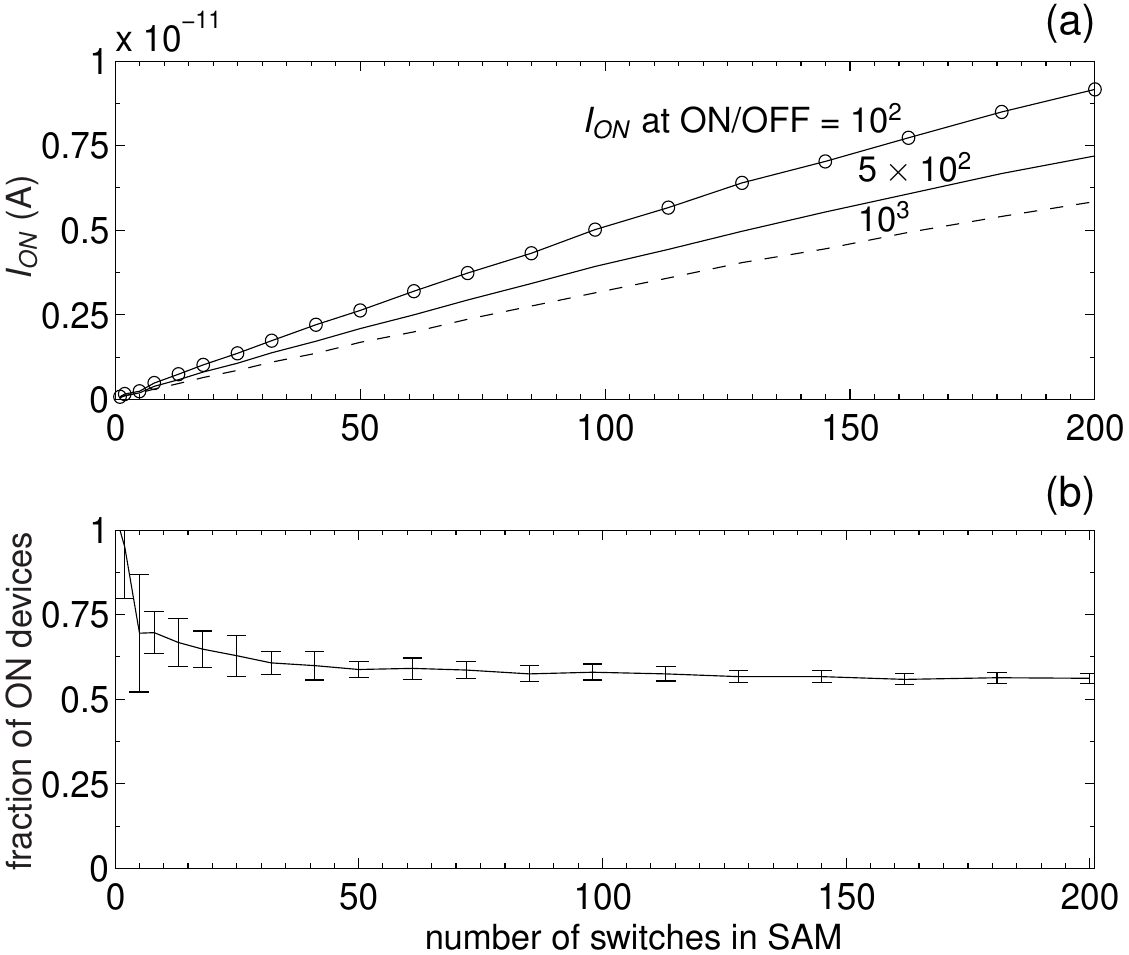}
\caption{Summary of Monte Carlo simulations of SAMs of various area: (a) the average ON currents at voltages providing certain ON/OFF ratios; (b) the average fraction of negatively charged traps at the equilibrating voltage $V_e$.}
\label{fig16}
\end{figure}

Figure \ref{fig15} shows the resulting average correlation between molecules (and its fluctuations) as a function of the distance between them in the $5 \times 5$-switch SAM. The charge states of neighboring traps are significantly anticorrelated, while the next-next neighbor charge states are positively correlated. This means that the switching is due to a nearly-simultaneous entry of electrons into roughly every other trap \footnote{there is virtually no correlation between the transistor molecules, just as with their autocorrelation in time \cite{korotkov94}, because at least two transition channels are open at any time.}. This explains why in the top inset in Fig. \ref{fig14} the average fraction of charged traps is close to 1/2. Thus the only adverse effect of the Coulomb interaction between individual resistive switches is the approximately two-fold reduction of the average ON current per device. Figure \ref{fig16} presents a summary characterization of the SAM operation as a function of its size (and hence its area).

\begin{figure}[!th]
\centering
\includegraphics[width=21pc]{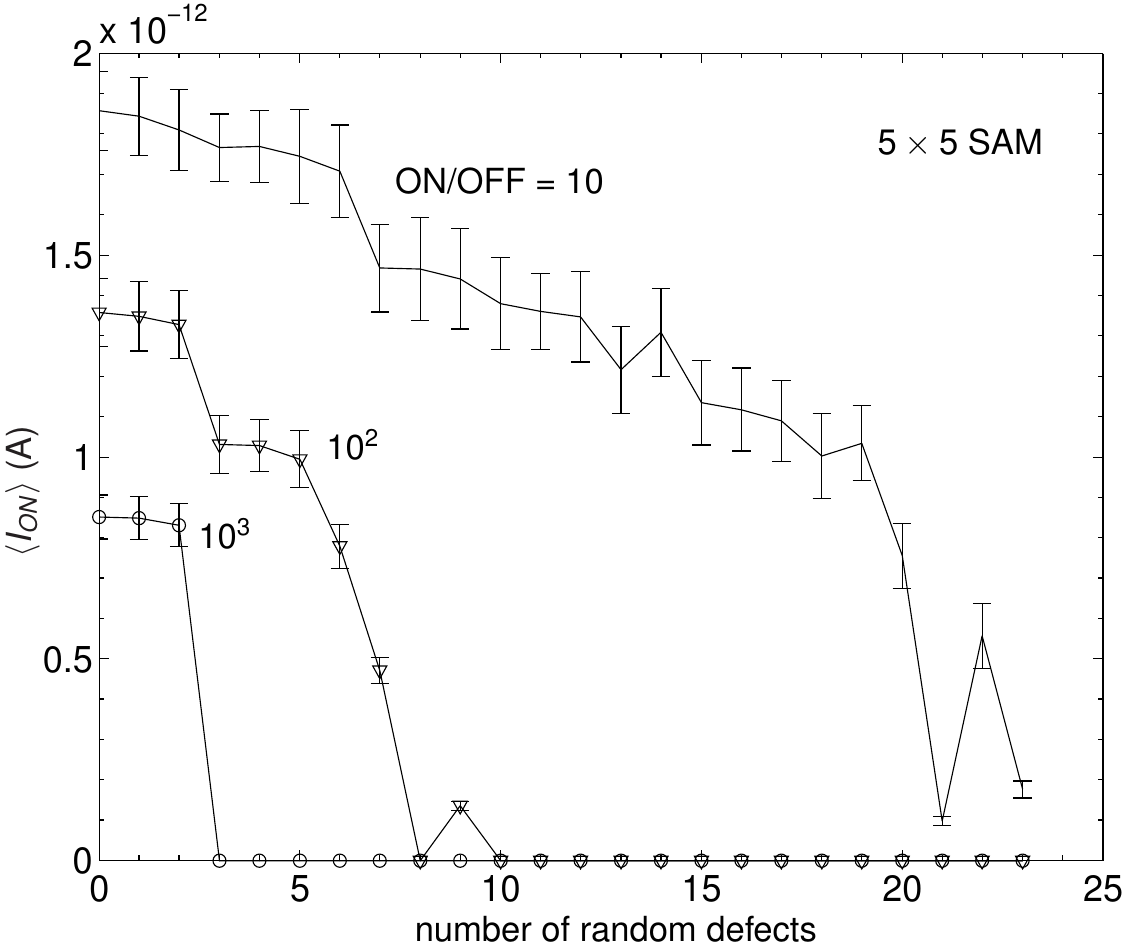}
\caption{Defect tolerance of the $5 \times 5$ SAM switch: ON current as a function of a number of molecules held artificially in a fixed, random charge state, at random locations, at the applied voltage values necessary to ensure a certain level of the ON/OFF current ratio. Error bars show the r.m.s. spread of results.}
\label{fig17}
\end{figure}

The fact that even the fractional charging of traps in SAMs is sufficient for a very good modulation of their net current suggests that these devices should have a high tolerance to defects and stray electric charges \cite{likharev99}. In order to verify this, we have carried out a preliminary evaluation of the defect tolerance by artificially fixing charge states of certain, randomly selected component molecules. The results, shown in Fig. \ref{fig17}, are rather encouraging, implying that the switches may provide the ON/OFF current ratios above 100 at defective switch fractions up to $\sim 10$\%, and at a comparable concentration of random offset charges.

\section{Conclusion}

Despite the problems with the description of single-electron charging in the density-functional theory, described in detail in Appendices A and B, we have managed to combine its advanced (ASIC) version to analyze the possibility of using single-electron tunneling effects in molecular assemblies for the implementation of bistable memristive devices (``resistive switches''). Our results indicate that chemically-plausible molecules and self-assembled monolayers of such molecules may indeed operate, at room temperature, as nonvolatile resistive switches which would combine multi-year retention times with sub-second switching times, and have ON/OFF current ratios in excess of $10^3$. Moreover, we have obtained strong evidence that operation of the SAM version of the device is tolerant to a rather high concentration of defects and randomly located charged impurities. The ON current of a single device ($\sim 0.1$ pA at $V \approx 2$ V) corresponds to a very reasonable density ($\sim 4$ W/cm$^2$) of the power dissipated in an open SAM switch, potentially enabling 3D integration of hybrid CMOS/nano circuits \cite{ChengStrukov12}. (Note that the average power density in a crossbar is at least 4 times lower because of the necessary crosspoint device spacing (Fig. \ref{fig1}d); besides that, in all applications we are aware of, at least 50\% of the switches (and frequently much more) are closed, decreasing the power even further.) 

However, even our best design (Fig. \ref{fig2}) still requires additional work. First, proper spatial positions of the functional molecules have to be enforced by some additional molecular support groups which have not been taken into account in our analysis yet. If the spacer groups fixing the relative spatial arrangement of the islands can be constructed from saturated molecular units similar to the alkane chains used to separate the islands from the electrodes, then the calculations presented here should be applicable to complete devices, but this expectation still has to be verified.

Second, we feel that there is room for improvement in the choice of molecular chains used as tunnel barriers and intermediate islands. For example, the low calculated effective mass, $m_{ef} \approx 0.1 m_0$, of electrons tunneling along alkane chains makes it necessary to use rather long chains, despite their large HOMO-LUMO gaps (which, in turn, require large switching voltages --- see Figs. \ref{fig4}, \ref{fig10}). The use of a molecular chain with a higher $m_{ef}$ and a narrower gap would decrease switching voltages (and hence energy dissipation at switching), and also reduce the total device length, resulting in shorter switching times (at the same charge retention).

Third, the defect tolerance of SAM-based switches should be evaluated in more detail, for charged impurities located not only on the molecular acceptor groups, but also between them --- say, inside the (still unspecified) support groups. 

Finally, an experimental verification of our predictions looks imperative for the further progress of work towards practicable molecular resistive switches.

\section*{Acknowledgment}

This work was supported by the Air Force Office of Scientific Research. The supercomputer resources used in this work were provided by DOD's HPCMP. Valuable comments by P. Allen, D. Averin and M. Fernandez-Serra are gratefully acknowledged.  We would also like to thank C. Pammaraju and S. Sanvito for their generous help with the ASIC SIESTA software package. 

\appendix
\section{Single-electron charging correction}

Let us consider a simple but reasonable model of a well-conducting (say, metallic) island, of a size well above the Thomas-Fermi screening length, in which the single-electron addition energies are simply

\begin{equation}
\Delta E(i) = K_i - e\phi_i,
 \label{eq_A1}
\end{equation}
where $K_i$ is $i$-th electron's kinetic energy (which, as well as the island capacitance $C$, is assumed to be independent of other electron state occupancies, but is an arbitrary function of $i$), and the second term describes the potential energy of that electron in the net electrostatic potential of all other charges,

\begin{equation}
\phi_i=\phi_0-(i-1)\frac{e}{C},
 \label{eq_A2}
\end{equation}
where $\phi_0$ is the background potential of the nuclei, and the second term is due to the previously added electrons. In this model the total ground-state energy of an $n$-electron ion (besides the electron-independent contributions) is

\begin{eqnarray}
E_{gr}(n)&=&\sum_{i=1}^n\Delta E(i) \nonumber \\
         &=&\sum_{i=1}^n K_i - en\phi_0+\frac{e^2}{2C}n(n-1),
   \label{eq_A3}
\end{eqnarray}
so that the energy difference created by the last charging is

\begin{eqnarray}
\Delta E(n)&=&E_{gr}(n)-E_{gr}(n-1) \nonumber \\
           &=&K_n-e\phi_0+\frac{e^2}{C}(n-1).
   \label{eq_A4}
\end{eqnarray}

On the other hand, in a hypothetical na\"ive DFT theory, without the partial self-interaction corrections present in its LSDA, GGA and ASIC versions, the single-particle (Kohn-Sham) energies of ion $n$ of this model are written as 

\begin{equation}
\varepsilon_i^{\mbox{\scriptsize DFT}}(n)=K_i-e\phi_n,\mbox{  }\phi_n=\phi_0-\frac{e}{C}n.
 \label{eq_A5}
\end{equation}
For the calculation of the full ground-state energy of ion $n$, such generic DFT sums up these energies from $i = 1$ to $i = n$, adding the ``double-counting correction'' term \cite{jones89}, in the Gaussian units equal to
\begin{equation}
E_{\mbox{\scriptsize corr}}=-\frac{1}{2}\int d^3r \int d^3r^\prime \frac{\rho(\textbf{r})\rho(\textbf{r}^\prime)}{\left|\textbf{r}-\textbf{r}^\prime \right|},
 \label{eq_A6}
\end{equation}
where $\rho(\textbf{r})$ is the total electron charge density at point \textbf{r}. For our simple model, this correction is just $–e^2n^2/2C$, so that

\begin{eqnarray}
E_{gr}^{\mbox{\scriptsize DFT}}&=&\sum_{i=1}^n\varepsilon_i^{\mbox{\scriptsize DFT}}(n)-\frac{e^2n^2}{2C}  \nonumber \\
 &=&\sum_{i=1}^nK_i-en\phi_0+\frac{e^2n^2}{2C},
   \label{eq_A7}
\end{eqnarray}
and

\begin{eqnarray}
\Delta E^{\mbox{\scriptsize DFT}}(n)& \equiv & E_{gr}^{\mbox{\scriptsize DFT}}(n)-E_{gr}^{\mbox{\scriptsize DFT}}(n-1) \nonumber \\
   &=& K_n-e\phi_0+\frac{e^2}{C}\left(n-\frac{1}{2}\right).
   \label{eq_A8}
\end{eqnarray}
Comparing this result with Eq. (\ref{eq_A4}), we obtain the following relation:

\begin{equation}
\Delta E(n)=\Delta E^{\mbox{\scriptsize DFT}}(n)-\frac{e^2}{2C}.
 \label{eq_A9}
\end{equation}

Thus in the na\"ive DFT theory, the single-electron addition energy differs from the correct expression  (\ref{eq_A4}) by $e^2/2C$.  Moreover, it does not satisfy the fundamental Eq. (\ref{eq_epsilon_def_ground}). Indeed, for $i = n$, Eq. (\ref{eq_A9}) gives the following result,

\begin{equation}
\varepsilon_n^{\mbox{\scriptsize DFT}}(n)=K_n-e\phi_0+\frac{e^2}{C}n
 \label{eq_A10}
\end{equation}
which, according to Eqs. (\ref{eq_A4}) and (\ref{eq_A8}) may be rewritten either as

\begin{equation}
\Delta E(n) = \varepsilon_n^{\mbox{\scriptsize DFT}}(n)-\frac{e^2}{C},
 \label{eq_A11}
\end{equation}
or as

\begin{equation}
\Delta E^{\mbox{\scriptsize DFT}}(n) = \varepsilon_n^{\mbox{\scriptsize DFT}}(n)-\frac{e^2}{2C}.
 \label{eq_A12}
\end{equation}
This error is natural, because such DFT version ignores the fundamental physical fact that an electron does not interact with itself, even if it is quantum-mechanically spread over a finite volume. This difference can become quite substantial in small objects such as molecular groups. For example, Table 1 shows the results using LSDA SIESTA calculations for two different ions of our trap molecule (Fig. \ref{fig2}), with $n = n_0 + 1$ and $n = n_0 +2$, where $n_0 = 330$ is the total number of protons in the molecule. The results show that the inconsistency described by Eq. (\ref{eq_A12}) is indeed very substantial and is independent (as it should be) of the applied voltage $V$ in the range keeping the working orbital's energy inside the HOMO-LUMO gap of the alkane chain. The two last columns of the tables show the values of $e^2/2C$, calculated in two different ways: from the relation following from Eq. (\ref{eq_A5}):

\begin{equation}
\frac{e^2}{2C}=\frac{\varepsilon_n^{\mbox{\scriptsize DFT}}(n)-\varepsilon_n^{\mbox{\scriptsize DFT}}(n-1)}{2},
 \label{eq_A13}
\end{equation}
and from the direct electrostatic expression

\begin{equation}
\frac{e^2}{2C}=\frac{1}{2}\int \phi_n(\textbf{r})\left|\psi_n^{\mbox{\scriptsize DFT}}(\textbf{r})\right|^2d^3r,
 \label{eq_A14}
\end{equation}
where $\phi_n(\textbf{r})$ is the part of the electrostatic potential, created by the electron of the $n$-th orbital of the $n$-th ion. The values are very close to each other and correspond to capacitance $C \approx 4.5 \times 10^{-20}$ F which a perfectly conducting sphere of diameter $d \approx 0.8$ nm would have. The last number is in a very reasonable correspondence with the size of the acceptor group of the molecule --- see Fig. \ref{fig2}.

\begin {table}[!th]
\caption {Columns 2 and 3: values of the single-electron transfer energy $\Delta E(n)$ for the trap molecule ions with $n = n_0 + 1$ and $n = n_0 + 2$ electrons, calculated in LSDA SIESTA and then self-interaction corrected as discussed in Appendix A, as functions of the applied voltage (Column 1). Columns 3 and 4 list the values of parameter $e^2/C$ , calculated as discussed in Appendix A.} 
\label{tab:tab1} 
\begin{flushleft}
\begin{tabular}{ >{\centering\arraybackslash}m{1.5cm}  >{\centering\arraybackslash}m{1.6cm} >{\centering\arraybackslash}m{1.6cm} >{\centering\arraybackslash}m{1.6cm} >{\centering\arraybackslash}m{1.6cm}}
\toprule[1.5pt]
{ Voltage $V$ (V)} & { $\Delta E(n)$ from Eq. (\ref{eq_A9})} (eV) & {  $\Delta E(n)$ from Eq. (\ref{eq_A11}) (eV) } & { $e^2/2C$ from Eq. (\ref{eq_A13}) (eV)} & { $e^2/2C$ from Eq. (\ref{eq_A14}) (eV)}\\
\hline
\midrule
\multicolumn{5}{c}{$n=n_0+1$} \\
\hline
\cmidrule(r){1-5}
-2.36 &	-3.08 &	-3.01 &	1.84 &	1.79 \\
-1.18 &	-3.38 &	-3.37 &	1.84 &	1.79 \\
0.00 &	-3.73$^{\mbox{\scriptsize (a)}}$ &	-3.73$^{\mbox{\scriptsize (a)}}$ &	1.84 &	1.79\\
1.18 &	-4.07 &	-4.10 &	1.84 &	1.79\\
2.36 &	-4.42 &	-4.46 &	1.84 &	1.79\\
3.53 &	-4.77 &	-4.82 &	1.84 &	1.79\\
\cmidrule(r){1-5}
\hline
\multicolumn{5}{c}{$n=n_0+2$} \\
\cmidrule(r){1-5}
\hline
7.07 &	-1.91 &	-2.02 &	1.82 &	1.79 \\
8.24 &	-2.29 &	-2.39 &	1.82 &	1.79 \\
9.42 &	-2.61 &	-2.75 &	1.82 &	1.79 \\
10.60 &	-2.97 &	-3.11 &	1.82 &	1.79 \\
11.78 &	-3.32 &	-3.47 &	1.82 &	1.79 \\

\bottomrule[1.25pt]
\end {tabular}
\\[1.5pt] 
$^{\mbox{\scriptsize (a)}}$ The numbers to be compared with experimental values of electron affinity: -3.31 eV Ref. \cite{bhozale08} and -3.57 eV Ref. \cite{singh06}.
\end{flushleft}
\end{table}

The second and third columns of the table present the genuine electron addition energies $\Delta E(n)$ calculated from, respectively, Eq. (\ref{eq_A9}) and (\ref{eq_A11}), using the average of the above values of $e^2/2C$. Not only do these values coincide very well; they are in a remarkable agreement with experimentally measured electron affinities \cite{singh06, bhozale08} of molecules similar to our molecular trap. 

We believe that these results show that, first, LSDA SIESTA provides very small compensation of the self-interaction effects in the key energy $\Delta E(n)$ and, second, that (at least for the lowest negative ions of our trap molecules), an effective compensation may be provided using any of the simple relations (\ref{eq_A9}) and (\ref{eq_A11}).

\section{Level freezing in DFT}

\begin{figure}[!th]
\centering
\includegraphics[width=21pc]{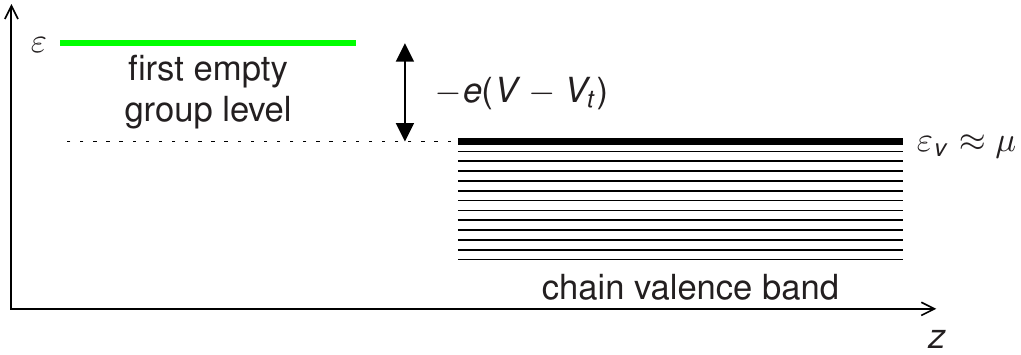}
\caption{The schematic energy spectrum of our model at a voltage $V$ below voltage $V_t$ that aligns the group localized level $\varepsilon$ with the valence band edge $\varepsilon_v$.}
\label{fig_B1}
\end{figure}

For the analysis of the fictitious ``level freezing'' predicted by a na\"ive DFT at $V > V_t$ (see Fig. \ref{fig6}), let us consider the following simple model: a molecule consisting of a small acceptor group with just one essential energy level, and a spatially separated chain with a quasi-continuous valence band. Figure \ref{fig_B1} shows the energy spectrum of the system at $V < V_t$. (As before, the occupied levels are shown in black, while the unoccupied ones are shown in green.)

The edge $\varepsilon_v$ of the band is separated from the first unoccupied level in the group by energy $-e(V - V_t)$, where $V$ is the fraction of the voltage drop between the centers of the group and the tail of a molecule, and $V_t$ is its value which aligns the level with $\varepsilon_v$.  Now let $V$ be close to $V_t$, so that the occupancy $p$ of the discrete level is noticeable. If the effect of group charging on the exchange-correlation energy is negligible, a generic DFT theory (e.g., LSDA) would describe the system energy as 

\begin{eqnarray}
E& = & E_0-e(V-V_t)p \nonumber \\
   &+&\frac{1}{2}\int d^3r \int d^3r^\prime \frac{\rho(\textbf{r})\rho(\textbf{r}^\prime)-\rho_0(\textbf{r})\rho_0(\textbf{r}^\prime)}{\left|\textbf{r}-\textbf{r}^\prime\right|},
   \label{eq_B1}
\end{eqnarray}
where index 0 marks the variable values at $p = 0$.

Now let us simplify Eq. (\ref{eq_B1}) by assuming that due to a small size of the acceptor group, the Coulomb interaction of electrons localized on it is much larger than that on the chain, so that the latter may be neglected. (For the trap molecule shown in Fig. 2, this assumption is true within $\sim$5\%.) Then Eq. (\ref{eq_B1}) is reduced to

\begin{eqnarray}
E &\approx& E_0 - e(V-V_t)p - \frac{e}{2}\int_{\mbox{\scriptsize group}} \phi(\textbf{r})\left|\psi(\textbf{r})\right|^2d^3r, \nonumber \\
p&=&\int_{\mbox{\scriptsize group}}\left|\psi(\textbf{r})\right|^2d^3r,
   \label{eq_B2}
\end{eqnarray}
where $\phi(\textbf{r})$ is the electrostatic potential created by the part of the electronic wavefunction that resides on the group. In the simple capacitive model of the group (used in particular in Appendix A), $\phi(\textbf{r}) = -ep/C$, where $C$ is the effective capacitance of the group, so that

\begin{equation}
E \approx E_0 - e(V-V_t)p + \frac{e^2p^2}{2C}.
 \label{eq_B3}
\end{equation}

On the other hand, in accordance with Eq. (\ref{eq_A7}), the total energy in the DFT may also be presented in the form

\begin{equation}
E= \sum_{i}p_i\varepsilon_i - \frac{1}{2}\int d^3r \int d^3r^\prime \frac{\rho(\textbf{r})\rho(\textbf{r}^\prime)}{\left|\textbf{r}-\textbf{r}^\prime\right|},
   \label{eq_B4}
\end{equation}
where $\varepsilon_i$ are all occupied (or partially occupied) single particle energies, so that in our simple model

\begin{equation}
E \approx E_0 + (\varepsilon-\varepsilon_v)p - \frac{e^2p^2}{2C}.
   \label{eq_B5}
\end{equation}
Comparing Eqs. (\ref{eq_B3}) and (\ref{eq_B5}), we arrive at the following expression:

\begin{equation}
\varepsilon-\varepsilon_v \approx -e(V-V_t) + \frac{e^2p}{C}.
   \label{eq_B6}
\end{equation}
In most DFT packages, level occupancies $p_i$ are calculated from the single-particle Fermi distribution,

\begin{equation}
p_i=\frac{1}{\mbox{exp}\left\{(\varepsilon_i-\mu)/k_{\mbox{\scriptsize B}}T\right\}+1};
   \label{eq_B7}
\end{equation}
for our simple model, index $i$ may be dropped, and (due to the valence band multiplicity) $\mu \approx \varepsilon_v$.

\begin{figure}[!th]
\centering
\includegraphics[width=21pc]{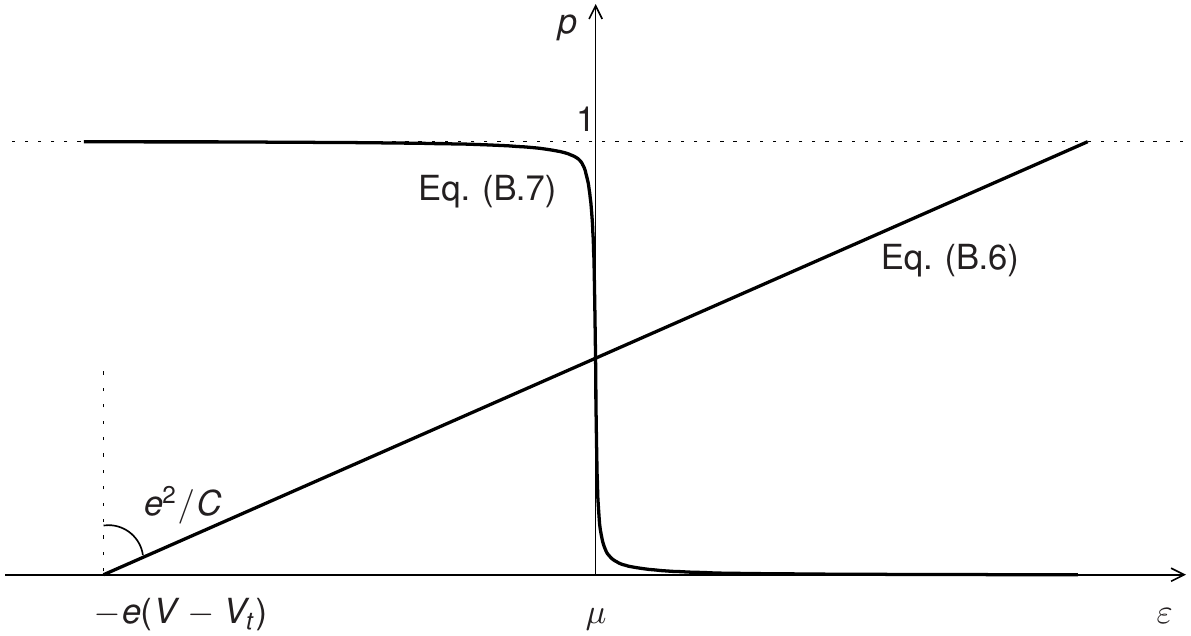}
\caption{A sketch of Eqs. (\ref{eq_B6}) and (\ref{eq_B7}).}
\label{fig_B2}
\end{figure}

As is evident from the sketch of Eqs. (\ref{eq_B6}) and (\ref{eq_B7}), in Fig. \ref{fig_B2}, if the thermal fluctuation scale $k_{\mbox{\scriptsize B}}T$ is much lower than the charging energy scale $e^2/C$, then almost within the whole range $V_t < V < V_t + e/C$, the approximate solution of the system of these equations is

\begin{equation}
p \approx \frac{C}{e}(V-V_t),\mbox{ }\varepsilon \approx \varepsilon_v.
   \label{eq_B8}
\end{equation}

\begin{figure}[!th]
\centering
\includegraphics[width=21pc]{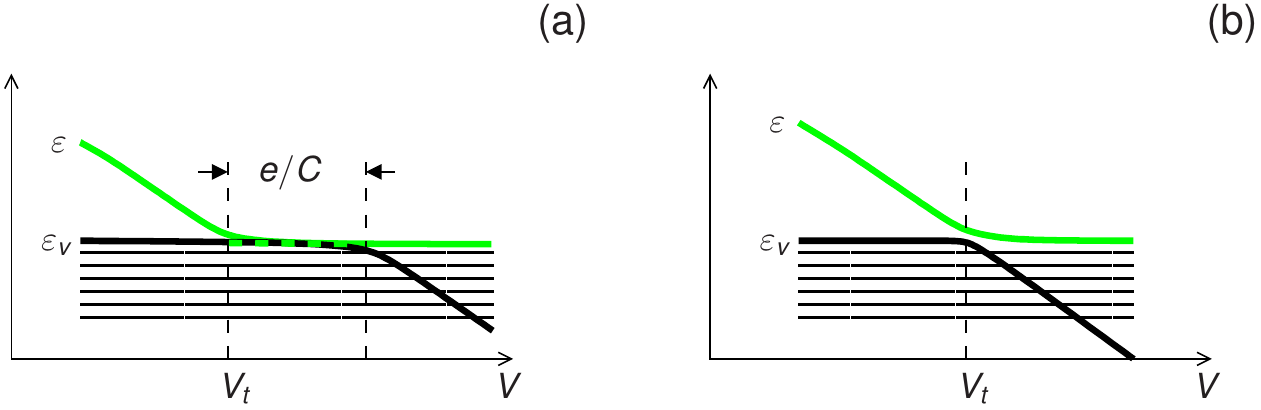}
\caption{(a) A sketch of the evolution of the energy spectrum from Fig. \ref{fig_B1} as a function of the applied voltage $V$, illustrating the self-interaction errors giving rise to a spurious level freezing in the $V_t < V < V_t + e/C$ voltage range. The dashed black-green line indicates the region with a partial occupancy $0 < p < 1$ of the group-localized orbital. (b) A sketch of the evolution of the same energy spectrum in a correct quantum-mechanical theory, in which electrons do not self-interact.}
\label{fig_B3}
\end{figure}

Panel (a) in Fig. \ref{fig_B3} shows (schematically) the resulting dependence of the energy spectrum of our model system on the applied voltage $V$, with level freezing in the range $V_t < V < V_t + e/C$. The dashed black-green line indicates the region with a partial occupancy $0 < p < 1$ of the group-localized orbital. In panel (b) in (Fig. \ref{fig_B3}) we show the evolution which should follow from the correct quantum-mechanical theory, in which electrons do not self-interact and as a result there is the usual anticrossing of energy levels $\varepsilon$ and $\varepsilon_v$ at $V = V_t$. (For clarity, Fig. \ref{fig_B3} strongly exaggerates the anticrossing width, which is less than $10^{-3}$ eV for our trap \footnote{Direct SIESTA calculation has shown that the anticrossing energy splitting is less than the calculation error (of the order of $10^{-3}$ eV). An indirect calculation using Eq. (\ref{eq_bardeen}), with $\psi_W$ and $\psi_v$ substituted instead of $\psi_{i}$ and $\psi_{s,c}$, suggests that this overlap is as small as $\sim 10^{-8}$ eV.}.)

The actual spectrum of our molecular trap is somewhat more complex than that of the simple model above --- see Figs. \ref{fig4}a and \ref{fig6}. First, not only the valence energy band of the alkane chain, but also its conduction band is important for electron transfer in our voltage range. Second, the molecular group has not one, but a series of discrete energy levels, with the most important of them corresponding to the working orbital (energy $\varepsilon_W$), and one more group-localized orbital with energy $\varepsilon_{W+1} \approx \varepsilon_W + 0.7$ eV.

\begin{figure}[!th]
\centering
\includegraphics[width=21pc]{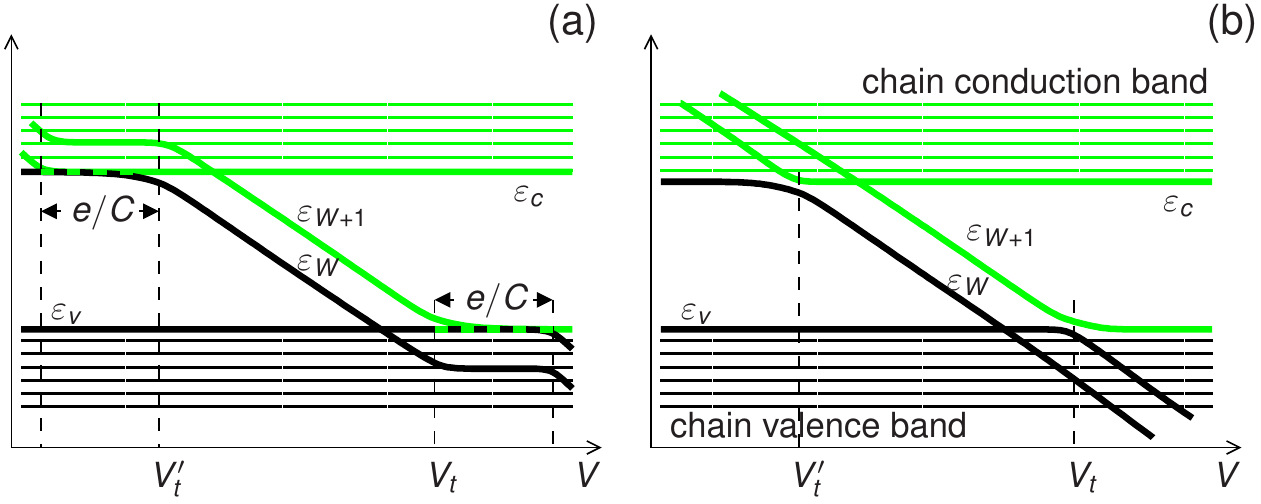}
\caption{(a) A sketch of the evolution of the molecular energy spectrum of our trap molecule as a function of the applied voltage $V$, illustrating the self-interaction errors giving rise to a spurious level freezing in voltage ranges $V_t^\prime - e/C < V < V_t^\prime$ and $V_t < V < V_t + e/C$. The dashed black-green line indicates the region with a partial occupancy $0 < p < 1$ of the group-localized orbitals with energies $\varepsilon_W$ or $\varepsilon_{W+1}$. (b) A sketch of the evolution of the molecular energy spectrum but in a correct quantum-mechanical theory, in which electrons do not self-interact.}
\label{fig_B4}
\end{figure}

\begin{figure}[!th]
\centering
\includegraphics[width=21pc]{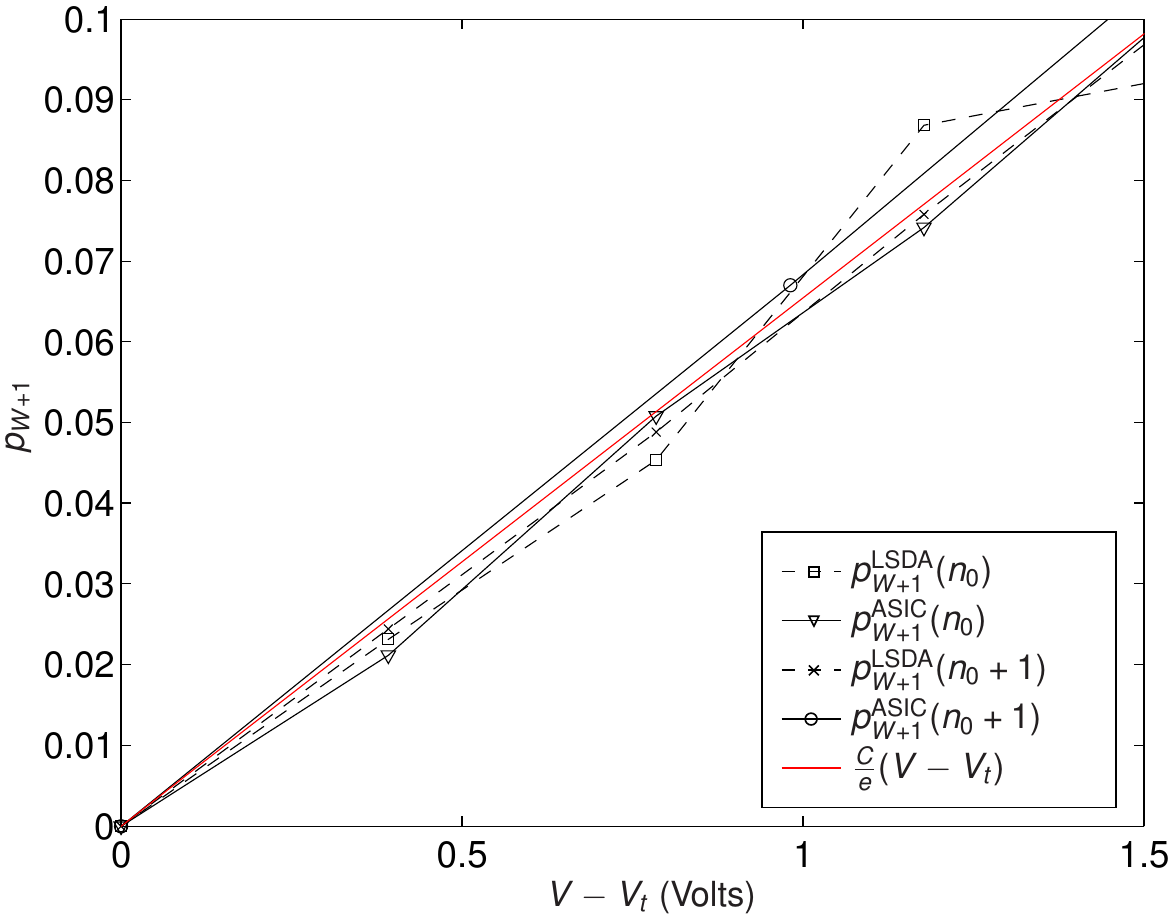}
\caption{The DFT-calculated occupancy $p_{W+1}$  of the $(W+1)$'st orbital of the acceptor group of our trap molecule at voltages above the threshold voltage $V_t$ of the alignment of its energy $\varepsilon_{W+1}$ with alkane chain's valence band edge $\varepsilon_v$. Black lines show results of two versions of DFT theory, for two ion states: the singly-negatively charged ion and the neutral molecule, while the red line shows the result given by Eq. (\ref{eq_B8}) with $C = 4.5\times 10^{-20}$ F.}
\label{fig_B5}
\end{figure}

Nevertheless, the behavior of the spectrum, predicted by uncorrected versions of DFT (Fig. \ref{fig6}) may still be well understood using our model. Just as was discussed above, for voltages $V$ above the threshold value $V_t$ (which now corresponds to the alignment of $\varepsilon_v$ with $\varepsilon_{W+1}$ rather than $\varepsilon_{W}$), it describes a gradual transfer of an electron between the top level of the valence band and the second group-localized orbital, with its occupation number $p_{W+1}$ gradually growing in accordance with Eq. (\ref{eq_B8}) --- see panel (a) in Fig. \ref{fig_B4}. Similarly, at voltages $V$ below $V_t^\prime$ (which corresponds to the alignment of the working orbital's energy $\varepsilon_W$ with the lowest level $\varepsilon_{c}$ of the chain's conduction band), there is a similar spurious gradual transfer of an electron between the corresponding orbitals. In both voltage ranges, a spurious internal electrostatic potential is created; as is described by Eq. (\ref{eq_B8}), it closely compensates the changes of the applied external potential, thus ``freezing'' all orbital energies of the system at their levels reached at thresholds $V_t^\prime$ and $V_t$ --- see panel (a) in Fig. \ref{fig_B4}. Figure \ref{fig_B5} shows that results of both the LSDA and ASIC DFT calculations at $V > V_t$ agree well with Eq. (\ref{eq_B8}), with a value $C = 4.5 \times 10^{-20}$ F calculated as discussed in Appendix A, indicating that the electron self-interaction effects remain almost uncompensated in these software packages, at least for complex molecules such as our trap.

Again, in the correct quantum-mechanical theory, there should be a simple (and in our molecules, extremely narrow) anticrossing between the effective single-particle levels of the acceptor group and the alkane chain --- see panel (b) in Fig. \ref{fig_B4}. As described in Sec. III of the main text, we have succeeded to describe this behavior rather closely, using the internal iteration dynamics of ASIC SIESTA with $T = 0$ K.

\bibliography{msbib}

\end{document}